\documentclass[twocolumn,aps,epsfig,nofootinbib]{revtex4}
\usepackage{graphicx,epstopdf,latexsym,amssymb,amsmath,color,mathrsfs,rotating}
\usepackage[center]{subfigure}

\begin{document}
\allowdisplaybreaks
 \newcommand{\bq}{\begin{equation}}
 \newcommand{\eq}{\end{equation}}
 \newcommand{\bqn}{\begin{eqnarray}}
 \newcommand{\eqn}{\end{eqnarray}}
 \newcommand{\nb}{\nonumber}
 \newcommand{\lb}{\label}
 \newcommand{\f}{\frac}
 \newcommand{\p}{\partial}
\newcommand{\PRL}{Phys. Rev. Lett.}
\newcommand{\PLB}{Phys. Lett. B}
\newcommand{\PRD}{Phys. Rev. D}
\newcommand{\CQG}{Class. Quantum Grav.}
\newcommand{\JCAP}{J. Cosmol. Astropart. Phys.}
\newcommand{\JHEP}{J. High. Energy. Phys.}

\title{Scalar and tensor perturbations in loop quantum cosmology: High-order corrections}
 
\author{Tao Zhu, Anzhong Wang\footnote{Corresponding author}}
\affiliation{Institute for Advanced Physics $\&$ Mathematics, Zhejiang University of Technology, Hangzhou, 310032, China\\
GCAP-CASPER, Physics Department, Baylor University, Waco, TX 76798-7316, USA}

\author{Gerald Cleaver}
\affiliation{EUCOS-CASPER, Physics Department, Baylor University, Waco, TX 76798-7316, USA}

\author{Klaus Kirsten and Qin Sheng}
\affiliation{GCAP-CASPER, Mathematics Department, Baylor University, Waco, TX 76798-7328, USA}

\author{Qiang Wu}
\affiliation{Institute for Advanced Physics $\&$ Mathematics, Zhejiang University of Technology, Hangzhou, 310032, China}

\date{\today}

\begin{abstract}

Loop quantum cosmology (LQC) provides promising resolutions to the trans-Planckian issue and initial singularity arising in the inflationary models of general relativity. In general, due to different quantization approaches, LQC involves two types of quantum corrections, the holonomy and inverse-volume, to both of the cosmological background evolution and perturbations. In this paper, using {\em the third-order  uniform asymptotic approximations}, we derive explicitly the observational quantities  of the slow-roll inflation in the framework of LQC with these quantum corrections. We calculate   the power spectra, spectral indices, and running of the spectral indices for both scalar and tensor perturbations, whereby the tensor-to-scalar ratio is obtained. We expand all the observables at the time when the inflationary mode crosses the Hubble horizon. As the upper error bounds for the uniform asymptotic approximation at the third-order are $\lesssim 0.15\%$, these results represent the most accurate results obtained so far in the literature. It is also shown that with the inverse-volume corrections, both scalar and tensor spectra exhibit a deviation from the usual shape at large scales.  Then, using the Planck, BAO and SN data we obtain new constraints on quantum gravitational effects from LQC corrections, and find that such effects could be within the detection of the  forthcoming experiments.

\end{abstract}


\maketitle

\section{Introduction}
\renewcommand{\theequation}{1.\arabic{equation}} \setcounter{equation}{0}

The inflationary cosmology provides the simplest and most elegant mechanism to produce the primordial density perturbations and primordial gravitational waves (PGWs) \cite{Guth, InfGR}. The former grows to produce the large-scale structure (LSS) seen today in the universe, and meanwhile creates the cosmic microwave background (CMB) temperature anisotropy, which has been extensively probed by WMAP \cite{WMAP}, PLANCK \cite{PLANCK}, and other CMB experiments as well as galaxy surveys. PGWs, on the other hand, produce not only a temperature anisotropy, but also a distinguishable signature in CMB polarization--the B-mode, which was once thought to be   already observed by BICEP2 \cite{BICEP2}, although subsequent analysis of multi-frequency data from BICEP2/Keck and Planck Collaborations showed that the signals could be  due to galactic dust \cite{Planck-intermediate} (see also \cite{Mort2014}), and further confirmation is needed. These observations in the measurement of the power spectra and spectral indices, together with the forthcoming ones, provide unique opportunities for us to gain deep insight into the physics of the very early Universe. More importantly, they may provide a unique window to explore quantum gravitational  effects, which otherwise cannot be studied in the near future by any man-made terrestrial experiments.

Inflation is very sensitive to Planckian physics \cite{DB}. In particular, in most inflationary scenarios, the energy scale of the inflationary fluctuations, which relates to the present observations, was not far from the Planck scale at the beginning of inflation. As in such high energy regime the usual classical general relativity and effective field theory are known to be broken, and it is widely expected a quantum theory of gravity could provide a complete description of the early Universe. However, such a theory of quantum gravity has not been established yet, and only a few candidates exist. One of the promising approaches is  loop quantum gravity. On the basis of this theory, loop quantum cosmology (LQC) was proposed, which offers a natural framework to address the trans-Planckian issue and initial singularity, arising in the inflation scenarios. In fact, in LQC, because of the quantum gravitational effects deep inside the Planck scale, the big bang singularity is replaced by a big bounce \cite{Bojowald2001, Ashtekar2006}. This remarkable feature has motivated a lot of interest to consider the underlying quantum geometry effects in the standard inflationary scenario and their detectability \cite{QGEs}.

In LQC, roughly speaking, there are two kinds of quantum gravitational corrections to the cosmological background and cosmological perturbations: the holonomy \cite{Mielczarek2008, Grain2009PRL, Grain2010PRD, vector_holonomy, scalar, scalar2, loop_corrections} and the inverse-volume \cite{Bojowald2008, Bojowald2009, Bojowald2007, Bojowald2008b, Bojowald2011, Bojowald2011b}.  The main consequence of holonomy corrections on the cosmological background is to replace the big bang singularity of the Friedmann-Lema\^itre-Robertson-Walker  (FLRW) universe by a big bounce. The cosmological scalar \cite{scalar}, vector \cite{vector_holonomy}, and tensor perturbations \cite{Mielczarek2008, Grain2009PRL, Grain2010PRD} have been calculated explicitly with holonomy corrections (see also \cite{scalar2}).  Due to these corrections, the modifications of the algebra of constraints generically leads to anomalies \cite{vector_holonomy, scalar}. Recently, it has been shown that these anomalies can be removed by adjusting the form of the quantum corrections to the Hamiltonian constraint. This is achieved by adding suitable counter terms that vanish in the classical limit \cite{vector_holonomy, scalar, scalar2}. With these anomaly-free cosmological perturbations, the dispersion relation of the inflationary mode function $\mu_k(\eta)$ (of scalar and tensor perturbations) is modified by the quantum corrections to the form \cite{scalar2, loop_corrections},
\bqn
\lb{omega_h}
\omega^2_{k}(\eta) = \left(1-2 \frac{\rho(\eta)}{\rho_c}\right)k^2,
\eqn
where $\rho_c$ is the energy density at which the big bounce happens. In the classical limit $\rho \ll \rho_c$, the above equation reduces to the standard one of GR. From this equation, one can see that when the energy density $\rho$ approaches ${\rho_c}/{2}$, the dynamics of cosmological perturbations is significantly modified by the holonomy corrections. With this modified dispersion relation, the power spectra for both scalar and gravitational wave perturbations were calculated up to the first-order approximations of the slow-roll parameters, and the observability of these corresponding quantum gravitational effects were discussed in some detail \cite{loop_corrections, Mielczarek2014}.

The inverse-volume corrections are due to terms in the Hamiltonian constraint which cannot be quantized directly but only after being re-expressed as a Poisson bracket. It was demonstrated that the algebra of cosmological scalar \cite{Bojowald2008, Bojowald2009}, vector \cite{Bojowald2007}, and tensor perturbations \cite{Bojowald2008b} with quantum corrections can be closed. Consequently, the cosmological perturbed equations for scalar and tensor perturbations are modified \cite{Bojowald2011,Bojowald2011b}. For the scalar perturbations, the dispersion relation  takes the form \cite{Bojowald2011}
\bqn\lb{omega_i}
\omega^{2}_{k}(\eta) =
\left\{1+ \left[\frac{\sigma \vartheta_0}{3} \left(\frac{\sigma}{6}+1\right)+\frac{\alpha_0}{2}\left(5-\frac{\sigma}{3}\right)
\right] \delta_{\text{Pl}}(\eta)\right\}k^2,\nb\\
\eqn
where the constants $\alpha_0$, $\vartheta_0$, and $\sigma$ encode the specific features of the model, and  $\delta_{\text{Pl}}(\eta)$ is time-dependent, which usually behaves like $\delta_{\text{Pl}} \sim a^{-\sigma}$. For tensor perturbations, $\omega^{2}_{k}(\eta)$ is given by \cite{Bojowald2011},
\bqn
\lb{eom5}
\omega^2_{k}(\eta) &=& \Big(1+2 \alpha_0 \delta_{\text{Pl}}(\eta)\Big) k^2.
\eqn
The power spectra of both scalar and tensor perturbations due to the inverse-volume corrections, again up to the first-order approximations of the slow-roll parameters,  were studied in \cite{Bojowald2011}, in which  some constraints on some parameters of the model were obtained  from observational data \cite{Bojowald2011b}. The non-Gaussianities with inverse-volume corrections was also discussed in \cite{Cai2012}.

Although a lot of effort has already been devoted to the inflationary models of LQC with both holonomy and inverse-volume quantum corrections,  very accurate calculations of inflationary observables in LQC are still absent, and with  the arrival of the era of precision cosmology, such calculations are highly demanded. Recently, we have developed a powerful method, {\em the uniform asymptotic approximation method} \cite{Zhu1, Zhu2, Uniform3, Uniform4}, to make precise observational predictions from inflation models, after quantum gravitational effects are taken into account. We note here that such method was first applied to inflationary cosmology in \cite{uniformPRL}, and then we have developed it to more general mult-turning points cases and more precise higher order approximations \cite{Zhu1, Zhu2, Uniform3, Uniform4}. The main purpose of the present paper is to use this powerful method to derive the inflationary observables in LQC with holonomy and inverse-volume quantum corrections with high accuracy. More specifically, we consider the slow-roll inflation with the quantum gravitational corrections, but ignore the pre-inflation dynamics. By using the general expressions of power spectra, spectral indices, and running of spectral indices obtained in \cite{Uniform3}, we calculate explicitly these quantities, up to the third-order approximations in terms of the uniform asymptotic approximation parameter, for which the upper error bounds are $\lesssim 0.15\%$. All the inflationary observables are expressed in terms of slow-roll parameters and parameters representing quantum corrections explicitly. In the present paper, we also provide the expansion of all these inflationary observables at the time when the inflationary mode crosses the Hubble horizon, and calculate the tensor-to-scalar ratio. It is interesting to note that the holonomy corrections do not contribute to the tensor-to-scalar ratio up to the third-order approximation. More interestingly, it is shown that with the inverse-volume corrections, both scalar and tensor spectra exhibit a deviation from the standard one at large scales, which could provide a smoking gun for further observations.

The paper is organized as follows. In Sec. II, we present all the background evolution and perturbation equations with the holonomy corrections, and calculate explicitly the power spectra, spectral indices, and running of spectral indices. Then,  in Sec. III we turn to consider both background and perturbations with the inverse-volume corrections, and calculate all the inflationary observables. Our main conclusions are summarized in Sec. IV. Three appendices are also included. In Appendix A, we give a brief introduction to the uniform asymptotic approximation method with high-order corrections, while in Appendices B and C, we present some quantities discussed in the content of the paper.

Part of the results to be presented in this paper was reported recently in \cite{Zhu3}. In this paper, we shall provide detailed derivations of these results, and meanwhile report our studies of other aspects of LQC inflationary cosmology.

\section{Inflationary observables with holonomy corrections}
\renewcommand{\theequation}{2.\arabic{equation}} \setcounter{equation}{0}
In this section, let us consider the inflationary cosmology with the  holonomy corrections.

\subsection{Background equations and equations of motion for scalar and tensor perturbations}

To begin with, let us first consider a flat FLRW background
\bqn
ds^2=a^2(\eta) (-d\eta^2+dx^i dx^i),
\eqn
where $a(\eta)$ is the expansion  factor and $\eta$ the conformal time. With the holonomy corrections, the Friedmann equation is modified to the form \cite{Ashtekar2006}
\bqn
H^2=\frac{8\pi G}{3} \rho \left(1-\f{\rho}{\rho_c}\right),
\eqn
where $H=\dot a/a$ is the Hubble parameter with a dot representing derivative with respect to the cosmic time $t$ ($dt \equiv ad\eta$), $\rho$ is the energy density of the matter content, and $\rho_c$ is a characteristic energy scale of the holonomy corrections and usually is of the order of the Planck energy density: $\rho_c \sim m_{\text{Pl}}^4$ with the Planck mass $m_{\text{Pl}}=1.22\times 10^{19} \text{GeV}$. A general prediction associated with the above equation is that the holonomy corrections lead to a resolution of the big bang singularity, in which it is replaced by a non-singular big bounce occurring at $\rho=\rho_c$. For a scalar field  $\varphi$ with potential $V(\varphi)$, the Klein-Gordon equation reads
\bqn
\ddot \varphi+3 H \dot \varphi+\f{dV(\varphi)}{d\varphi}=0.
\eqn
The energy density of the inflaton field is
\bqn
\rho=\f{\dot \varphi^2}{2}+V(\varphi).
\eqn

The above set of equations determines uniquely the evolution of the FLRW background. As shown in  \cite{slow-roll-holonomy}, ``the standard slow-roll inflation" can be triggered by the preceding phase of quantum bounce. In this paper, for the sake of simplification, we shall focus on the slow-roll inflation with the holonomy corrections, and ignore the pre-inflation dynamics. In this case, we are in a process such that the energy density of the cosmological fluid is supposed to be dominated by the potential of inflaton $\varphi$, i.e., $\dot \varphi^2 \ll V(\varphi)$. With this condition, it is convenient to define a hierarchy of Hubble flow parameters,
\bqn\lb{epsilon}
\epsilon_0 \equiv \f{H_{\text{ini}}}{H},\;\;\;\epsilon_{n+1}\equiv \f{d\ln \epsilon_n}{d\ln a}.
\eqn
On the other hand, as the quantum holonomy correction is also small, it is convenient to introduce the parameter $\delta_H$ by,
\bqn\lb{deltaH}
\delta_H \equiv \frac{\rho}{\rho_c} \ll 1.
\eqn
Then we have $\Omega\equiv 1- 2 \rho/\rho_c=1-2 \delta_H $.

Let us now consider cosmological perturbations. With the holonomy corrections, the equations for cosmological scalar perturbations can be cast in a modified Mukhanov equation for the gauge-invariant mode function  $\mu_k^{(s)}(\eta)$ \cite{scalar, scalar2},
\bqn\lb{eom-scalar}
\f{d^2\mu_k^{(s)}(\eta)}{d\eta^2}+\left(\omega_k^2(\eta)-\frac{z_s''(\eta)}{z_s(\eta)}\right)\mu_k^{(s)}(\eta)=0,
\eqn
where $\omega_k^2(\eta)=\Omega(\eta) k^2$, and $z_s \equiv a \dot \varphi /H$. Similarly,  the equation for the tensor perturbation  can be cast in the form  \cite{scalar2},
\bqn\lb{eom-tensor}
\f{d^2\mu_k^{(t)}(\eta)}{d\eta^2}+\left(\Omega(\eta)k^2-\frac{z_t''(\eta)}{z_t(\eta)}\right)\mu_k^{(t)}(\eta)=0,
\eqn
where $z_t\equiv a/\sqrt{\Omega}$.

\subsection{Power spectra and spectral indices in the uniform asymptotic approximation}

To apply for the uniform asymptotic approximation, first we write the equations of motion for both scalar (Eq.\ (\ref{eom-scalar})) and tensor (Eq.\ (\ref{eom-tensor}))
perturbations   in the form (\ref{eom}) by introducing a new variable $y=-k\eta$. Then, we find that $\hat g(y)$ and $q(y)$ must be chosen as \cite{Zhu1,Zhu2},
\bqn
\lambda^2 \hat g(y)&=&-\frac{1}{k^2} \left(\Omega(\eta)-\frac{z''(\eta)}{z(\eta)}\right)+\frac{1}{4y^2}\nb\\
&=& \frac{\nu^2(\eta)}{y^2}-c_s^2(\eta),\\
q(y)&=& - \f{1}{4y^2},
\eqn
where $c_s(\eta) = \sqrt{\Omega(\eta)}$ and $\nu^2(\eta)=\eta^2 z''(\eta)/z(\eta)+1/4$. For the scalar perturbations, using the slow-roll parameters defined in Eqs. (\ref{epsilon}) and (\ref{deltaH}), we find that
\bqn\lb{zsh}
\frac{z_s''(\eta)}{z_s(\eta)}&\simeq&a^2H^2\Big(2-\epsilon_{1}+\frac{3 \epsilon _2}{2} +\frac{\epsilon _2^2}{4}+\frac{\epsilon _2 \epsilon _3}{2}-\frac{\epsilon _1 \epsilon _2}{2}\nb\\
&&~~~-6 \epsilon _1 \delta _H+6 \epsilon _1^2 \delta _H-4 \epsilon _1 \epsilon _2 \delta _H-18 \epsilon _1 \delta _H^2\Big).\nb\\
\eqn
In the above, we used the subscript (or superscript) ``s" to denote quantities associated with the scalar perturbations. Similarly,  for the tensor perturbations, we find
\bqn\lb{zth}
\frac{z_t''(\eta)}{z_t(\eta)} &\simeq &a^2 H^2 \Big(2-\epsilon _1-6 \delta _H \epsilon_1-18 \delta _H^2 \epsilon _1-48 \delta _H^3 \epsilon_1\nb\\
&&~~+6 \delta _H \epsilon _1^2+38 \delta _H^2 \epsilon _1^2-2 \delta _H \epsilon _1 \epsilon _2-6 \delta _H^2 \epsilon _1 \epsilon _2\Big),\nb\\
\eqn
where the subscript (or superscript) ``t" denotes quantities associated with the tensor perturbations.

With the functions $\hat g(y)$ and $q(y)$ given above, we are in the position to calculate the power spectra and spectral indices from the general formulas Eq.\ (\ref{pw}) and Eq. (\ref{indices}). As we discussed in \cite{Zhu2}, in order to do so, first we need to expand $\nu(\eta)$ and $c_s(\eta)$ around the turning point $\bar y_0$ of $\hat g(y)$ (i.e., $\hat g(\bar y_0)=0$ with $\bar y_0=-k \eta_0$), then perform the integral of $\sqrt{\hat g(y)}$ in Eq.\ (\ref{pw}), and calculate the error control function $\mathscr{H}(+\infty)$. In the slow-roll inflation, it is convenient to consider the following expansions,
\bqn
\nu(\eta) &\simeq& \bar \nu_0 + \bar \nu_1 \ln\f{y}{\bar y_0}+\frac{1}{2} \bar \nu_2 \ln^2\f{y}{\bar y_0},\\
c_s(\eta) &\simeq& \bar c_0  + \bar c_1 \ln\f{y}{\bar y_0}+\frac{1}{2} \bar c_2 \ln^2\f{y}{\bar y_0},
\eqn
with
\bqn
\bar \nu_1 \equiv \f{d\nu(\eta)}{d\ln(-\eta)}\Bigg|_{\eta_0},\;\;\;\;\;\bar \nu_2 \equiv \f {d^2 \nu(\eta)}{d\ln^2(-\eta)} \Bigg|_{\eta_0},\nb\\
\bar c_1 \equiv \f{dc_s(\eta)}{d\ln(-\eta)}\Bigg|_{\eta_0},\;\;\;\;\;\bar c_2 \equiv \f {d^2 c_s(\eta)}{d\ln^2(-\eta)} \Bigg|_{\eta_0}.
\eqn
In the above, $\eta_0$ is the conformal time for mode $k$ at the turning point $\bar y_0$. In the slow-roll inflation with the holonomy corrections, in general we have $\nu(\eta) \simeq \frac{3}{2}+\mathcal{O}(\epsilon)$, $\bar \nu_1 \simeq \mathcal{O}(\epsilon^2)$, $\bar \nu_2 \simeq \mathcal{O}(\epsilon^3)$, and $\bar c_0 \simeq 1+\mathcal{O}(\delta_H)$, $\bar c_1 \simeq \mathcal{O}(\epsilon \delta_H)$, $\bar c_2 \simeq \mathcal{O}(\epsilon^2 \delta_H)$. The slow-roll expansions of all these quantities are presented in Appendix C.

With the above expansions, we notice that $\sqrt{ g(y)}=\sqrt{\lambda^2 \hat g(y)}$ can be expanded as
\bqn\lb{ge}
\sqrt{g(y)} &\simeq& \frac{\sqrt{\bar \nu_0^2-\bar c_0^2 y^2}}{y}+\frac{\bar \nu_0 \bar \nu_1-\bar c_0 \bar c_1 y^2}{y \sqrt{\bar \nu_0^2-\bar c_0^2 y^2}}\ln\frac{y}{\bar y_0}\nb\\
&&+\Bigg(\frac{\bar \nu_0 \bar \nu_2 }{2y \sqrt{\bar \nu_0^2-\bar c_0^2 y^2}}-\frac{\bar c_0 \bar c_2 y}{2 \sqrt{\bar \nu_0^2-\bar c_0^2 y^2}}\nb\\
&&\;\;\;\;\;\;\;\;\;-\frac{(\bar c_0 \bar \nu_1+\bar \nu_0 \bar c_1)^2y}{2 (\bar \nu_0^2-\bar c_0^2 y^2)^{3/2}}\Bigg)\ln^2\frac{y}{\bar y_0}.
\eqn
Therefore, the integral $\int \sqrt{ g}dy$ is divided into three parts,
\bqn\lb{g3parts}
\int_{y}^{\bar y_0} \sqrt{ g(\hat y)}dy=I_1+I_2+I_3,
\eqn
where
\bqn
\lim_{y\to 0}I_1&=& -\bar \nu_0 \left(1+\ln \frac{y}{2\bar y_0}\right),\nb\\
\lim_{y\to 0}I_2&=&\frac{(1-\ln2) \bar c_1 \bar \nu_0}{\bar c_0}-\left(\frac{ \pi^2}{24}-\frac{\ln^22}{2}+\frac{1}{2}\ln^2\frac{y}{\bar y_0}\right)\bar \nu_1,\nb\\
\lim_{y\to 0}I_3&=&- \bar \nu_0 \left(\frac{\pi^2-12\ln^22}{24}\right) \left(\frac{\bar c_1}{\bar c_0}-\frac{\bar \nu_1}{\bar \nu_0}\right)^2\nb\\
&&-\bar \nu_0 \left(1-\frac{\pi^2}{24}-\ln2+\frac{\ln^22}{2}\right)\frac{\bar c_2}{\bar c_0}\nb\\
&&+\left(\frac{\zeta(3)}{4}-\frac{\pi^2\ln2}{24}+\frac{\ln^23}{6}-\frac{1}{6}\ln^3\frac{y}{\bar y_0}\right)\bar \nu_2.\nb\\
\eqn
In the above, $\zeta(n)$ denotes the Riemann zeta function.

Now, we turn to consider the error control function $\mathscr{H}$, which in general can be written as
\bqn
\mathscr{H}(\xi)&=&\frac{5}{36} \left\{\int_{\bar y_0}^{\tilde{y}}\sqrt{\hat{g}(y')}dy'\right\}^{-1}\Bigg|^{y}_{\bar y_0}\nb\\
&&-\int_{\bar y_0}^{y} \left\{\frac{q}{\hat{g}}-\frac{5\hat{g}'^2}{16\hat{g}^3}+\frac{\hat{g}''}{4\hat{g}^2}\right\}\sqrt{\hat{g}}dy.
\eqn
In the limit $y\to 0$, after some lengthy calculations, the above expression can be cast in the form
\bqn\lb{Hinfty}
\frac{\mathscr{H}(+\infty)}{\lambda}&\simeq& \frac{1}{6\bar \nu_0} \left(1+\frac{\bar c_1}{\bar c_0}\right)-\frac{\bar\nu_1(23+12\ln 2)}{72\bar \nu_0^2}.\nb\\
\eqn
As we only need to calculate the power spectra up to the second-order in slow-roll parameters, in the above we have ignored $\bar c_1^2, \bar c_2, \bar \nu_1^2, \bar \nu_2$, terms. Once we get the integral of $\sqrt{ g(y)}$ (as presented in Eq.~(\ref{g3parts})) and error control function $\mathscr{H}(+\infty)$ (as presented in Eq.~(\ref{Hinfty})), from Eq.~(\ref{pw}) we can calculate the power spectra.

To calculate  the corresponding spectral indices, let us first consider the $k$-dependence of $\bar\nu_0(\eta_0)$, $\bar \nu_1(\eta_0)$ through  $\eta_0 = \eta_0(k)$. From the relation $-k\eta_0=\bar \nu_0(\eta_0)/\bar c_0(\eta_0)$, we observe that
\bqn
\frac{d\ln(-\eta_0)}{d\ln k}& \simeq& -1+\frac{\bar c_1}{\bar c_0}-\frac{\bar\nu_1}{\bar\nu_0}-\left(\frac{\bar c_1}{\bar c_0}-\frac{\bar\nu_1}{\bar\nu_0}\right)^2. ~~~
\eqn
Then, the spectral index is given by
\bqn
n-1&\simeq&
\left(3-2 \bar \nu _0\right)
+\frac{2 \bar c_1 \bar \nu_0}{\bar c_0}
+\left(\frac{1}{6\bar \nu_0^2}-2\ln2\right)\bar \nu_1
\nb\\
&&
+\left(-\frac{2 \bar \nu _0}{\bar c_0}-\frac{1}{6 \bar c_0 \bar \nu _0}
+\frac{2\bar  \nu _0 \ln2}{\bar c_0}\right)\bar c_2
\nb\\
&&
+\left(\frac{23+12\ln2}{72 \bar \nu _0^2}+\frac{\pi ^2}{12}-\ln^22\right)\bar \nu_2.
\eqn

Similarly, after some tedious calculations, we find that the running of the spectral index $\alpha \equiv dn/d\ln k$ can be written in the form
\bqn
\alpha(k)
&\simeq&
\frac{2 \bar \nu _0 \bar c_1^2}{\bar c_0^2}-\frac{4 \bar \nu _1 \bar c_1}{\bar c_0}+\left(\frac{1}{3 \bar \nu _0^3}+\frac{2}{\bar \nu _0}\right) \bar \nu _1^2\nb\\
&&+\left(-\frac{2 \ln2 \bar \nu _0}{\bar c_0}+\frac{2 \bar \nu _0}{\bar c_0}+\frac{1}{6 \bar c_0 \bar \nu _0}\right) \bar c_3+2 \bar \nu _1\nb\\
&&+\left(\ln4-\frac{1}{6 \bar \nu _0^2}\right) \bar \nu _2-\frac{2 \bar c_2 \bar \nu _0}{\bar c_0}\nb\\
&&+\left(\ln^22-\frac{\pi ^2}{12}-\frac{\ln2}{6 \bar \nu _0^2}-\frac{23}{72 \bar \nu _0^2}\right) \bar \nu _3.
\eqn

\subsection{Scalar perturbations}

With the slow-roll expansions of $\bar \nu_0,\;\bar \nu_1,\;\bar\nu_2,\;\bar\nu_3$ and $\bar c_1,\;\bar c_2,\;\bar c_3$ presented in the above and after some tedious calculations, we obtain the scalar spectrum,
\bqn
\Delta_s^2(k) &\simeq&\bar A_s \Bigg\{1+\bar \delta_H-2 \left(\bar D_{\text{p}}+1\right) \bar \epsilon _1-\bar D_{\text{p}} \bar \epsilon _2+\frac{3 }{2}\bar \delta _H^2\nb\\
&&~~~~-2\left(2 \bar D_{\text{p}}+3\right) \bar \delta _H \bar \epsilon _1-\bar D_{\text{p}} \bar \delta _H \bar \epsilon _2\nb\\
&&~~~~+\left(2 \bar D_{\text{p}}+2 \bar D_{\text{p}}^2+\frac{\pi^2}{2}-5+\bar \Delta_1 \right) \bar \epsilon _1^2\nb\\
&& ~~~~+\left(\bar D_{\text{p}}^2-\bar D_{\text{p}}+\frac{7\pi ^2}{12}-8+\bar \Delta_1+2 \bar \Delta_2 \right) \bar \epsilon _1 \bar \epsilon _2\nb\\
&&~~~~+\left(\frac{1}{2}\bar D_{\text{p}}^2+\frac{\pi^2}{8}-\frac{3}{2}+\frac{\bar \Delta_1}{4} \right) \bar \epsilon _2^2\nb\\
&&~~~~+\left(\frac{\pi ^2}{24}+\bar \Delta _1-\frac{\bar D_{\text{p}}^2}{2}\right) \bar \epsilon _2 \bar \epsilon _3\Bigg\},\nb\\
\eqn
where $\bar A_s \equiv \frac{181\bar H^2}{72 e^3\pi^2 \bar \epsilon_1}$, $\bar D_{\text{p}}\equiv \frac{67}{181}-\ln2$, $\bar \Delta_1 \equiv \frac{183606}{32761}-\frac{\pi^2}{2}$, and $\bar \Delta_2 \equiv \frac{9269}{589698}$. Note that a letter with an over bar  denotes a quantity evaluated at the turning point $\bar y_0$. Then, the scalar spectral index is
\bqn
n_s&\simeq&1-2 \bar \epsilon _1-\bar \epsilon _2+4 \bar \delta _H \bar \epsilon _1-2 \bar \epsilon _1^2-\bar D_{\text{n}}\bar \epsilon _2 \bar \epsilon _3\nb\\
&&-\left(2 \bar D_{\text{n}}+3\right) \bar \epsilon _1 \bar \epsilon _2+12 \bar \delta _H^2 \bar \epsilon _1+4 \bar D_{\text{n}}\bar \delta _H \bar \epsilon _1^2-2 \bar \epsilon _1^3\nb\\
&&-2 \left(\bar D_{\text{n}}-1\right) \bar \delta _H\bar \epsilon _1 \bar \epsilon _2\nb\\
&&-\left(6 \bar D_{\text{n}}-2\bar \Delta_{\text{n}1}-\pi^2+\frac{53}{3}\right) \bar \epsilon _1^2 \bar \epsilon_2\nb\\
&&-\left(3 \bar D_{\text{n}}+\bar D_{\text{n}}^2-\frac{7\pi ^2}{12}-\bar \Delta_{\text{n}1}-2\bar \Delta_{\text{n}2}+\frac{25}{3}\right) \bar \epsilon _1 \bar \epsilon _2^2\nb\\
&&+\left(\frac{\bar \Delta_{\text{n}1}}{2}+\frac{\pi^2}{4}-\frac{8}{3}\right) \bar \epsilon _2^2 \bar \epsilon_3\nb\\
&&-\left(4 \bar D_{\text{n}}+\bar D_{\text{n}}^2-\frac{7\pi ^2}{12}-\bar \Delta_{\text{n}1}-2\bar \Delta_{\text{n}2}+\frac{22}{3}\right) \bar \epsilon_1 \bar \epsilon _2 \bar \epsilon _3\nb\\
&&+\left(\bar \Delta_{\text{n}2}-\frac{\bar D_{\text{n}}^2}{2}+\frac{\pi ^2}{24}\right) (\bar \epsilon _2 \bar \epsilon _3^2+\bar \epsilon _2 \bar \epsilon _3 \bar \epsilon _4),\nb\\
\eqn
where $\bar{D}_{\text{n}}\equiv \frac{10}{27}-\ln2$, $\bar \Delta_{\text{n}1} \equiv \frac{454}{81}-\frac{\pi^2}{2}$, and $\bar \Delta_{\text{n}2} \equiv \frac{371}{2916}$. The running of the scalar spectral index reads
\bqn
\alpha_s &\simeq &
-2 \bar \epsilon _1 \bar \epsilon _2-\bar \epsilon _2 \bar \epsilon _3+4\bar \epsilon_1^2 \bar \delta_H-2 \bar \epsilon _1 \bar \epsilon _2 \bar \delta _H-6 \bar \epsilon _1^2 \bar \epsilon _2\nb\\
&&-\left(3+2 \bar D_{\text{n}}\right) \bar \epsilon _1 \bar \epsilon _2^2-2\left( \bar D_{\text{n}}+2\right) \bar \epsilon _1 \bar \epsilon _2 \bar \epsilon _3-\bar D_{\text{n}} \bar \epsilon _2 \bar \epsilon _3^2\nb\\
&&-\bar D_{\text{n}} \bar \epsilon _2 \bar \epsilon _3 \bar \epsilon _4-8 \bar D_{\text{n}}\bar \epsilon _1^3 \bar \delta _H-12 \bar \epsilon _1^3 \bar \epsilon _2\nb\\
&&+\left(2 \pi ^2-14 \bar D_{\text{n}}+4 \bar \Delta_{\text{n}1}-39\right) \bar \epsilon _1^2 \bar \epsilon _2^2\nb\\
&&+\left(\frac{7 \pi ^2}{12}-3 \bar D_{\text{n}}-\bar D_{\text{n}}^2+\bar \Delta_{\text{n}1}+2 \bar \Delta_{\text{n}2}-\frac{25}{3}\right) \bar \epsilon _1 \bar \epsilon _2^3\nb\\
&&+\left(\pi ^2-8 \bar D_{\text{n}}+2 \bar \Delta_{\text{n}1}-\frac{65}{3}\right) \bar \epsilon _1^2 \bar \epsilon _2 \bar \epsilon _3\nb\\
&&+\left(\frac{7 \pi ^2}{4}-10 \bar D_{\text{n}}-3 \bar D_{\text{n}}^2\right.\nb\\
&&~~~~~\left.+3 \bar \Delta_{\text{n}1}+6 \bar \Delta_{\text{n}2}-\frac{71}{3}\right) \bar \epsilon _1 \bar \epsilon _2^2 \bar \epsilon _3\nb\\
&&+\left(\frac{7 \pi ^2}{12}-5 \bar D_{\text{n}}-\bar D_{\text{n}}^2+\bar \Delta_{\text{n}1}+2 \bar \Delta_{\text{n}2}-\frac{22}{3}\right) \bar \epsilon _1 \bar \epsilon _2 \bar \epsilon _3^2\nb\\
&&+\left(\frac{\pi^2 }{2}+\bar \Delta_{\text{n}1}-5\right) \bar \epsilon _2^2 \bar \epsilon _3^2+\left(\frac{\pi ^2}{24}-\frac{\bar D_{\text{n}}^2}{2}+\bar \Delta_{\text{n}2}\right) \bar \epsilon _2 \bar \epsilon _3^3\nb\\
&&+\left(\frac{7 \pi ^2}{12}-5 \bar D_{\text{n}}-\bar D_{\text{n}}^2+\bar \Delta_{\text{n}1}+2 \bar \Delta_{\text{n}2}-\frac{22}{3}\right) \bar \epsilon _1 \bar \epsilon _2 \bar \epsilon _3 \bar \epsilon _4\nb\\
&&+\left(\frac{\pi }{4}+\frac{\bar \Delta_{\text{n}1}}{2}-\frac{8}{3}\right) \bar \epsilon _2^2 \bar \epsilon _3 \bar \epsilon _4\nb\\
&&+\left(\frac{\pi ^2}{8}-\frac{3 \bar D_{\text{n}}^2}{2}+3 \bar \Delta_{\text{n}2}\right) \bar \epsilon _2 \bar \epsilon _3^2 \bar \epsilon _4\nb\\
&&+\left(\frac{\pi ^2}{24}-\frac{\bar D_{\text{n}}^2}{2}+\bar \Delta_{\text{n}2}\right) \bar \epsilon _2 \bar \epsilon _3 \bar \epsilon _4^2+6\left(2 \bar D_\text{n}+3\right)  \bar \epsilon _1^2 \bar \epsilon _2 \bar \delta _H\nb\\
&&+\left(\frac{\pi ^2}{24}-\frac{\bar D_{\text{n}}^2}{2}+\bar \Delta_{\text{n}2}\right) \bar \epsilon _2 \bar \epsilon _3 \bar \epsilon _4 \bar \epsilon _5-2 \left(\bar D_{\text{n}}+3\right)\bar \epsilon _1 \bar \epsilon _2^2  \bar \delta _H\nb\\
&&-2 \left(\bar{D}_{\text{n}}+2\right)\bar \epsilon _1 \bar \epsilon _2 \bar \epsilon _3 \bar\delta _H+16\bar \epsilon_1^2 \bar \delta_H^2-6 \bar \epsilon_1\bar \epsilon_2 \bar \delta_H^2.\nb\\
\eqn

\subsection{Tensor perturbations}
Similar to the scalar perturbations, within the slow-roll approximations, we find that the power spectrum for the tensor perturbations reads
\bqn
\Delta_t^2(k) & \simeq &
\bar A_t
\Big\{1+\bar \delta_H-2 (\bar D_{\text{p}}+1) \bar \epsilon _1\nb\\
&&~~~~+\frac{3 \bar \delta_H^2}{2}-2 (2 \bar D_{\text{p}}+3) \bar \delta_H \bar \epsilon _1\nb\\
&&~~~~+\left(\bar \Delta _1+\frac{\pi ^2}{2}-5+2 \bar D_{\text{p}}+2 \bar D_{\text{p}}^2\right) \bar \epsilon _1^2\nb\\
&&~~~~+\left(2 \bar \Delta _2-2-2 \bar D_{\text{p}}-\bar D_{\text{p}}^2+\frac{\pi ^2}{12}\right) \bar \epsilon _1 \bar \epsilon _2\Big\},\nb\\
\eqn
where $\bar A_t \equiv \frac{181 \bar H^2}{36e^3 \pi^2}$. Also the tensor spectral index and its running are given by
\bqn
n_t & \simeq & -2 \bar \epsilon _1+4 \bar \delta _H \bar \epsilon _1-2 \bar \epsilon _1^2-2 (\bar{D}_{\text{n}}+1) \bar \epsilon _1\bar \epsilon _2+12 \bar \delta _H^2 \bar \epsilon _1\nb\\
&&+4 \bar D_{\text{n}} \bar \delta _H \bar \epsilon _1^2-2 \bar \epsilon _1^3-2 (\bar D_{\text{n}}-1) \bar \delta _H \bar \epsilon _1 \bar \epsilon _2\nb\\
&&+\left(\pi ^2-6 \bar D_{\text{n}}+2 \bar \Delta _{\text{n}1}-\frac{50}{3}\right) \bar \epsilon _1^2 \bar \epsilon _2\nb\\
&&+\left(\frac{\pi ^2}{12}-2 \bar D_{\text{n}}-\bar D_{\text{n}}^2+2 \bar \Delta _{\text{n}2}-2\right) \bar \epsilon _1 \bar \epsilon _2^2\nb\\
&&+\left(\frac{\pi ^2}{12}-2 \bar D_{\text{n}}-\bar D_{\text{n}}^2+2 \bar \Delta _{\text{n}2}-2\right) \bar \epsilon _1 \bar \epsilon _2 \bar \epsilon _3,\nb\\
\eqn
and
\bqn
\alpha_t &\simeq &
-2 \bar \epsilon _1 \bar \epsilon _2+4\bar  \delta _H \bar \epsilon _1^2-2\bar  \delta _H \bar \epsilon _1 \bar \epsilon _2-6 \bar \epsilon _1^2 \bar \epsilon _2-2 (\bar{D}_{\text{n}}+1) \bar \epsilon _1 \bar \epsilon _2^2\nb\\
&&-2 (\bar D_{\text{n}}+1) \bar \epsilon _1 \bar \epsilon _2 \bar \epsilon _3-8 \bar{D}_{\text{n}}\bar  \delta _H \bar \epsilon _1^3+2 (7+6 \bar{D}_{\text{n}})\bar  \delta _H \bar \epsilon _1^2 \bar \epsilon _2\nb\\
&&-2 (\bar{D}_{\text{n}}+2)\bar  \delta _H \bar \epsilon _1 \bar \epsilon _2^2-2 (\bar{D}_{\text{n}}+2)\bar  \delta _H \bar \epsilon _1 \bar \epsilon _2 \bar \epsilon _3-12 \bar \epsilon _1^3 \bar \epsilon _2\nb\\
&&+\left(2 \pi ^2-36-14 \bar{D}_{\text{n}}+4 \bar \Delta _{\text{n}1}\right) \bar \epsilon _1^2 \bar \epsilon _2^2\nb\\
&&+\left(\frac{\pi ^2}{12}-2 \bar D_{\text{n}}-\bar D_{\text{n}}^2+2 \bar \Delta _{\text{n}2}-2\right) \bar \epsilon _1 \bar \epsilon _2^3\nb\\
&&+\left(\pi ^2-8 \bar D_{\text{n}}+2 \bar \Delta _{\text{n}1}-\frac{56}{3}\right) \bar \epsilon _1^2 \bar \epsilon _2 \bar \epsilon _3\nb\\
&&+\left(\frac{\pi ^2}{4}-6 \bar D_{\text{n}}-3 \bar D_{\text{n}}^2+6 \bar \Delta _{\text{n}2}-6\right) \bar \epsilon _1 \bar \epsilon _2^2 \bar \epsilon _3\nb\\
&&+\left(\frac{\pi ^2}{12}-2 \bar D_{\text{n}}-\bar D_{\text{n}}^2+2 \bar \Delta _{\text{n}2}-2\right) \bar \epsilon _1 \bar \epsilon _2 \bar \epsilon _3^2\nb\\
&&+\left(\frac{\pi ^2}{12}-2 \bar D_{\text{n}}-\bar D_{\text{n}}^2+2 \bar \Delta _{\text{n}2}-2\right) \bar \epsilon _1 \bar \epsilon _2 \bar \epsilon _3 \bar \epsilon _4.
\eqn

 \subsection{Expansions at horizon crossing}
In the last two subsections, we have obtained the expressions of the power spectra, spectral indices, and running of spectral indices for both scalar and tensor perturbations. It should be noted that all these expressions were evaluated at the turning point $\bar y_0$. However, in the usual treatments, all expressions were expanded at the horizon crossing $a(\eta_\star) H(\eta_\star)= \sqrt{\Omega (\eta_\star)} k$. Thus, it is useful to rewrite all the expressions given in the last section at the time when the scalar (or tensor) mode crosses the horizon. After some tedious calculations, for the scalar perturbations, we find the scalar spectrum can be expressed as
 \bqn
 \Delta_s^2(k) &\simeq &
 A^{\star}_{s} \Big[ 1-2\left(1+D^{\star}_{\text{p}} \right) \epsilon _{\text{$\star $1}}-D^{\star}_{\text{p}} \epsilon _{\text{$\star $2}}+\delta _{\text{$\star $H}}\nb\\
 &&~~~~+\left(2D^{\star 2}_{\text{p}}+2D^{\star}_{\text{p}}+\frac{\pi^2}{2}-5+\Delta^{\star}_1\right) \epsilon _{\text{$\star $1}}^2\nb\\
 &&~~~~+\left(\frac{1}{2}D^{\star 2}_{\text{p}}+\frac{\pi^2}{8}-1+\frac{\Delta_1^{\star}}{4}\right) \epsilon _{\text{$\star $2}}^2\nb\\
&&~~~~+\frac{3}{2} \delta _{\text{$\star $H}}^2-D^{\star}_{\text{p}}\delta _{\text{$\star $H}} \epsilon _{\text{$\star $2}}-\left(4 D^{\star}_{\text{p}}+6\right) \delta _{\text{$\star $H}} \epsilon _{\text{$\star $1}}\nb\\
&&~~~~+\left(D^{\star 2}_{\text{p}}-D^{\star}_{\text{p}}+\frac{7\pi ^2}{12}-7+\Delta^{\star}_1+2\Delta^{\star}_2\right) \epsilon _{\text{$\star $1}} \epsilon _{\text{$\star $2}}\nb\\
&&~~~~+\left(\frac{\pi ^2}{24}-\frac{1}{2} D^{\star 2}_{\text{p}}+\Delta^{\star}_2\right) \epsilon _{\text{$\star $2}} \epsilon _{\text{$\star $3}}\Big],
\eqn
where the subscript ``$\star$" denotes evaluation at the horizon crossing, $A^{\star}_{s}\equiv \frac{181 H_\star^{2}}{72 e^3 \pi^2 \epsilon_{\star 1}}$, $D_{\text{p}}^{\star}=\frac{67}{181}-\ln3$, $\Delta_1^{\star} =\frac{485296}{98283}-\frac{\pi^2}{2}$, and $\Delta_2^{\star} =\frac{9269}{589698}$. For the scalar spectral index, one obtains
 \bqn
n_s&\simeq&
1-2 \epsilon_{\star 1}-\epsilon_{\star 2}-2 \epsilon_{\star 1}^2-(3+2 D_{\text{n}}^\star) \epsilon_{\star 1} \epsilon_{\star 2}\nb\\
&&-D_{\text{n}}^\star \epsilon_{\star 2} \epsilon_{\star 3}+4 \delta _{\text{$\star $H}} \epsilon _{\text{$\star $1}}+12 \delta _{\text{$\star $H}}^2 \epsilon _{\text{$\star $1}}\nb\\
&&+\left(-\frac{55}{3}-6 D_{\text{n}}^\star+\pi ^2+2 \Delta^{\star}_{\text{n}1}\right) \epsilon_{\star 1}^2 \epsilon_{\star 2}-2 \epsilon_{\star 1}^3\nb\\
&&+\left(\frac{7 \pi ^2}{12}+\Delta^{\star}_{\text{n}1}+2 \Delta^{\star}_{\text{n}2}-\frac{23}{3}-3 D_{\text{n}}^\star-D_{\text{n}}^{\star 2}\right) \epsilon_{\star 1} \epsilon_{\star 2}^2\nb\\
&&+\left(\frac{7 \pi ^2}{12}+\Delta^{\star}_{\text{n}1}+2 \Delta^{\star}_{\text{n}2}-\frac{23}{3}-4 D_{\text{n}}^\star-D_{\text{n}}^{\star 2}\right) \epsilon_{\star 1} \epsilon_{\star 2} \epsilon_{\star 3}\nb\\
&&+\left(\frac{\pi ^2}{4}+\frac{\Delta^{\star}_{\text{n}1}}{2}-\frac{7}{3}\right) \epsilon_{\star 2}^2 \epsilon_{\star 3}+\left(12\bar D_{\text{n}}-8D_{\text{n}}^\star\right) \delta _{\text{$\star $H}} \epsilon _{\text{$\star $1}}^2\nb\\
&&+\left(\frac{\pi ^2}{24}+\Delta^{\star}_{\text{n}2}-\frac{D_{\text{n}}^{\star 2}}{2}\right) ( \epsilon_{\star 2} \epsilon_{\star 3}^2+\epsilon_{\star 2} \epsilon_{\star 3} \epsilon_{\star 4})\nb\\
&&+\left(4D_{\text{n}}^\star-6 \bar D_{\text{n}}+2\right) \delta _{\text{$\star $H}} \epsilon _{\text{$\star $1}} \epsilon _{\text{$\star $2}}.
\eqn
The running of the scalar spectral index reads
\bqn
\alpha_s &\simeq&
-2 \epsilon_{\star1} \epsilon_{\star2}-\epsilon_{\star2} \epsilon_{\star3}-2(D_{\text{n}}^\star+2) \epsilon_{\star1} \epsilon_{\star2} \epsilon_{\star3}\nb\\
&&-6 \epsilon_{\star1}^2 \epsilon_{\star2}-D_\text{n}^\star \epsilon_{\star2} \epsilon_{\star3} \epsilon_{\star4}-(3+2D_{\text{n}}^\star) \epsilon_{\star1} \epsilon_{\star2}^2\nb\\
&&-D_\text{n}^\star \epsilon_{\star2} \epsilon_{\star3}^2+4 \delta _{\text{$\star $H}} \epsilon _{\text{$\star $1}}^2-2 \delta _{\text{$\star $H}} \epsilon _{\text{$\star $1}} \epsilon _{\text{$\star $2}}-12 \epsilon_{\star1}^3 \epsilon_{\star2}\nb\\
&&+\left(2 \pi ^2+4 \Delta^{\star} _{\text{n}1}-\frac{119}{3}-14D_{\text{n}}^\star\right) \epsilon_{\star1}^2 \epsilon_{\star2}^2\nb\\
&&+\left(\frac{7 \pi ^2}{12}+\Delta^{\star} _{\text{n}1}+2 \Delta^{\star} _{\text{n}2}\right.\nb\\
&&~~~~~~\left.-\frac{23}{3}-3D_{\text{n}}^\star-D_\text{n}^{\star2}\right) \epsilon_{\star1} \epsilon_{\star2}^3\nb\\
&&+\left(\pi ^2+2 \Delta^{\star}_{\text{n}1}-\frac{67}{3}-8D_{\text{n}}^\star\right) \epsilon_{\star1}^2 \epsilon_{\star2} \epsilon_{\star3}\nb\\
&&+\left(\frac{7 \pi ^2}{4}+3 \Delta^{\star} _{\text{n}1}+6 \Delta^{\star} _{\text{n}2}-23\right.\nb\\
&&~~~~~~-10D_{\text{n}}^\star-3D_{\text{n}}^{\star 2}\Bigg) \epsilon_{\star1} \epsilon_{\star2}^2 \epsilon_{\star3}-6 \delta _{\text{$\star $H}}^2 \epsilon _{\text{$\star $1}} \epsilon _{\text{$\star $2}}\nb\\
&&+\left(\frac{7 \pi ^2}{12}+\Delta^{\star} _{\text{n}1}+2 \Delta^{\star} _{\text{n}2}-5D_{\text{n}}^\star-D_\text{n}^{\star2}\right.\nb\\
&&~~~~~~~\left.-\frac{23}{3}\right) \epsilon_{\star1} \epsilon_{\star2} \epsilon_{\star3}^2\nb\\
&&+\left(\Delta^{\star} _{\text{n}1}+\frac{\pi ^2}{2}-\frac{14}{3}\right) \epsilon_{\star2}^2 \epsilon_{\star3}^2\nb\\
&&+\left(\Delta^{\star} _{\text{n}2}-\frac{D_\text{n}^{\star2}}{2}+\frac{\pi ^2}{24}\right) \epsilon_{\star2} \epsilon_{\star3}^3\nb\\
&&+\left(\frac{7 \pi ^2}{12}+\Delta^{\star} _{\text{n}1}+2 \Delta^{\star} _{\text{n}2}\right.\nb\\
&&~~~~~~\left.-\frac{23}{3}-5D_{\text{n}}^\star-D_\text{n}^{\star2}\right) \epsilon_{\star1} \epsilon_{\star2} \epsilon_{\star3} \epsilon_{\star4}\nb\\
&&+\left(\frac{\Delta^{\star} _{\text{n}1}}{2}+\frac{\pi ^2}{4}-\frac{7}{3}\right) \epsilon_{\star2}^2 \epsilon_{\star3} \epsilon_{\star4}+16 \delta _{\text{$\star $H}}^2 \epsilon _{\text{$\star $1}}^2\nb\\
&&+\left(3 \Delta^{\star} _{\text{n}2}-\frac{3}{2}D_{\text{n}}^{\star 2}+\frac{\pi ^2}{8}\right) \epsilon_{\star2} \epsilon_{\star3}^2 \epsilon_{\star4}\nb\\
&&+\left(\Delta^{\star} _{\text{n}2}-\frac{1}{2}D_\text{n}^{\star2}+\frac{\pi ^2}{24}\right) (\epsilon_{\star2} \epsilon_{\star3} \epsilon_{\star4}^2+\epsilon_{\star2} \epsilon_{\star3} \epsilon_{\star4} \epsilon_{\star5})\nb\\
&&-2\left(D_{\text{n}}^\star+3\right) \delta _{\text{$\star $H}} \epsilon _{\text{$\star $1}} \epsilon _{\text{$\star $2}}^2-2\left(D_{\text{n}}^\star+2\right) \delta _{\text{$\star $H}} \epsilon _{\text{$\star $1}} \epsilon _{\text{$\star $2}} \epsilon _{\text{$\star $3}}\nb\\
&&-8D_{\text{n}}^\star\delta _{\text{$\star $H}} \epsilon _{\text{$\star $1}}^3+6\left(2D_{\text{n}}^\star+3\right) \delta _{\text{$\star $H}} \epsilon _{\text{$\star $1}}^2 \epsilon _{\text{$\star $2}}.
\eqn

Similar to the scalar perturbations, now let us turn to consider the tensor perturbations, which yield
\bqn
\Delta_t^2(k) & \simeq &
A_t^\star \Bigg\{1-2\left(1+D_{\text{p}}^{\star}\right) \epsilon _{\text{$\star $1}}+\delta _{\text{$\star $H}}+\frac{3}{2}\delta _{\text{$\star $H}}^2\nb\\
&&~~~~~+\left(2D_{\text{p}}^{\star 2}+2D_{\text{p}}^{\star}+\frac{\pi^2}{2}-5+\Delta_1^{\star}\right) \epsilon _{\text{$\star $1}}^2\nb\\
&&~~~~~~+\left(-D_{\text{p}}^{\star 2}-2D_{\text{p}}^{\star}+\frac{\pi ^2}{12}-2+2\Delta_2^{\star}\right) \epsilon _{\text{$\star $1}} \epsilon _{\text{$\star $2}}\nb\\
&&~~~~~-2\left(2D^\star_{\text{p}}+3\right) \delta _{\text{$\star $H}} \epsilon _{\text{$\star $1}}\Bigg\}.
\eqn
 For the tensor spectral index, we find
\bqn
n_t &\simeq&
-2 \epsilon_{\star1}-2(1+D_{n}^\star) \epsilon_{\star1} \epsilon_{\star2}-2 \epsilon_{\star1}^2+4 \delta _{\text{$\star $H}} \epsilon _{\text{$\star $1}}\nb\\
&&-2 \epsilon_{\star1}^3+\left(\pi ^2-6 D_{n}^\star+2 \Delta^{\star} _{\text{n}1}-\frac{52}{3}\right) \epsilon_{\star1}^2 \epsilon_{\star2}\nb\\
&&+\left(\frac{\pi ^2}{12}-2+2 \Delta^{\star} _{\text{n}2}-2 D_{n}^\star-D_{n}^{\star2}\right) \epsilon_{\star1} \epsilon_{\star2}^2\nb\\
&&+\left(\frac{\pi ^2}{12}-2+2 \Delta^{\star} _{\text{n}2}-2 D_{n}^\star-D_{n}^{\star2}\right) \epsilon_{\star1} \epsilon_{\star2} \epsilon_{\star3}\nb\\
&&+\left(12\bar D_{\text{n}}-8 D_{\text{n}}^\star\right) \delta _{\text{$\star $H}} \epsilon _{\text{$\star $1}}^2\nb\\
&&+\left(4D_{\text{n}}^\star-6\bar D_{\text{n}}+2\right) \delta _{\text{$\star $H}} \epsilon _{\text{$\star $1}} \epsilon _{\text{$\star $2}}+12 \delta _{\text{$\star $H}}^2 \epsilon _{\text{$\star $1}}.\nb\\
\eqn
Then, the running of the tensor spectral index reads
\bqn
\alpha_t &\simeq&
-2 \epsilon _{\star1} \epsilon _{\star2}-2(1+D_{\text{n}}^\star) \epsilon _{\star1} \epsilon _{\star2} \epsilon _{\star3}-6 \epsilon _{\star1}^2 \epsilon _{\star2}\nb\\
&&-12 \epsilon _{\star1}^3 \epsilon _{\star2}-2(1+D_{\text{n}}^\star) \epsilon _{\star1} \epsilon _{\star2}^2+4 \delta _{\text{$\star $H}} \epsilon _{\text{$\star $1}}^2\nb\\
&&-2 \delta _{\text{$\star $H}} \epsilon _{\text{$\star $1}} \epsilon _{\text{$\star $2}}+16 \delta _{\text{$\star $H}}^2 \epsilon _{\text{$\star $1}}^2-8D_{\text{n}}^\star\delta _{\text{$\star $H}} \epsilon _{\text{$\star $1}}^3\nb\\
&&+\left(2 \pi ^2-14D_{\text{n}}^\star+4 \Delta^{\star} _{\text{n}1}-\frac{80}{3}\right) \epsilon _{\star1}^2 \epsilon _{\star2}^2\nb\\
&&+\left(\frac{\pi ^2}{12}+2 \Delta^{\star} _{\text{n2}}-2-2D_{\text{n}}^\star-D_\text{n}^{\star2}\right) \epsilon _{\star1} \epsilon _{\star2}^3\nb\\
&&+\left(\pi ^2-8D_{\text{n}}^\star+2 \Delta^{\star} _{\text{n}1}-\frac{58}{3}\right) \epsilon _{\star1}^2 \epsilon _{\star2} \epsilon _{\star3}\nb\\
&&+\left(\frac{\pi ^2}{4}-6+6 \Delta^{\star} _{\text{n2}}-6D_{\text{n}}^\star-3D_{\text{n}}^{\star2}\right) \epsilon _{\star1} \epsilon _{\star2}^2 \epsilon _{\star3}\nb\\
&&+\left(\frac{\pi ^2}{12}-2+2 \Delta^{\star} _{\text{n}2}-2D_{\text{n}}^\star-D_\text{n}^{\star2}\right) \epsilon _{\star1} \epsilon _{\star2} \epsilon _{\star3}^2\nb\\
&&+\left(\frac{\pi ^2}{12}-2+2 \Delta^{\star} _{\text{n}2}-2D_{\text{n}}^\star-D_\text{n}^{\star2}\right) \epsilon _{\star1} \epsilon _{\star2} \epsilon _{\star3} \epsilon _{\star4}\nb\\
&&-6 \delta _{\text{$\star $H}}^2 \epsilon _{\text{$\star $1}} \epsilon _{\text{$\star $2}}+2\left(6D_{\text{n}}^\star+7\right) \delta _{\text{$\star $H}} \epsilon _{\text{$\star $1}}^2 \epsilon _{\text{$\star $2}}\nb\\
&&-2\left(D_{\text{n}}^\star+2\right) \delta _{\text{$\star $H}} \epsilon _{\text{$\star $1}} \epsilon _{\text{$\star $2}}^2-2\left(D_{\text{n}}^\star+2\right) \delta _{\text{$\star $H}} \epsilon _{\text{$\star $1}} \epsilon _{\text{$\star $2}} \epsilon _{\text{$\star $3}}.\nb\\
\eqn

Finally with both scalar and tensor spectra given above, we can evaluate the tensor-to-scalar ratio at the horizon crossing time $\eta_\star$, and find that
\bqn
r &\simeq&
16 \epsilon _{\text{$\star $1}} \Bigg\{1+D_\text{p}^{\star} \epsilon _{\star2}+\left(\frac{17}{3}-\Delta _1^\star-\frac{\pi ^2}{2}+D_\text{p}^{\star}\right) \epsilon _1 \epsilon _2\nb\\
&&~~~~~~~+\left(\frac{7}{6}-\frac{\Delta _1^\star}{4}+\frac{D_\text{p}^{\star2}}{2}+\frac{D_\text{p}^{\star2}}{2}\right) \epsilon _{\star2}^2\nb\\
&&~~~~~~~+\left(\frac{D_\text{p}^{\star2}}{2}-\Delta _2^\star-\frac{\pi ^2}{24}\right) \epsilon _{\star2} \epsilon _{\star 3}\Bigg\}.\nb\\
\eqn
It is remarkable to note that the holonomy correction parameter $\delta_{\star H}$ doesn't contribute to the tensor-to-scalar ratio $r$, up to the third-order uniform asymptotic approximation.

 In addition, to the first-order of the slow-roll parameters, it can be shown that our results given above are consistent with those presented in \cite{Mielczarek2014}.

\section{Inflationary observables with inverse-volume corrections}
\renewcommand{\theequation}{3.\arabic{equation}} \setcounter{equation}{0}

Now let us turn to consider another type of quantum gravitational correction, the inverse-volume, in LQC.

\subsection{Background evolution and equations for perturbations}

In the presence of the inverse-volume corrections, the effective Friedmann and Klein-Gordon equations read \cite{Bojowald2011}
\bqn\lb{friedmann}
&& H^2=\frac{8\pi G}{3} \alpha \left(\frac{\dot \varphi^2}{2 \vartheta }+V(\varphi)\right),\\
\lb{KG}
&& \ddot \varphi +H \left(3-2\frac{d\ln \vartheta}{d\ln p}\right) \dot \varphi +\vartheta \frac{dV(\varphi)}{d\varphi}=0,
\eqn
with $p\equiv a^2$ and
\bqn\lb{alpha}
\alpha \simeq 1+\alpha_0 \delta_{\text{Pl}},\;\;\vartheta\simeq 1+\vartheta_0 \delta_{\text{Pl}},
\eqn
where $\delta_{\text{Pl}}$ characterizes the inverse-volume corrections in loop quantum cosmology and
\bqn
\delta_{\text{Pl}} \equiv \left(\frac{a_{\text{Pl}}}{a}\right)^\sigma.
\eqn
Note that here we only consider the inverse-volume correction $\delta_{\text{Pl}}$ at the first-order $\mathcal{O}(\delta_{\text{Pl}})$. Thus to be consistent, through the whole paper, we shall expand all the quantities at the first-order of $\delta_{\text{Pl}}$. In the above, $\alpha_0$, $\vartheta_0$, and $a_{\text{Pl}}$ are constants and depend on the specific models and parametrization of the loop quantization. Specifically for the parameter $\sigma$, different parametrization schemes shall provide different ranges of $\sigma$ \cite{minisuper,Bojowald2011}.  Moreover, $\alpha_0$ and $\vartheta_0$ are related by the consistency condition
\bqn
\vartheta_0 (\sigma-3)(\sigma+6)=3 \alpha_0 (\sigma-6),
\eqn
while $\sigma$ takes values in the range $0<\sigma\leq 6$.

The evolution of the background, which can be determined by the above set of equations, is usually different from the evolution given in the standard slow-roll inflation, because of the purely geometric effects of the inverse-volume corrections. However, as indicated in \cite{Bojowald2011b}, in that regime, the constraint algebra has not been shown to be closed. One way to consider the slow-roll inflation with inverse-volume corrections is in the large-volume regime, where the quantum corrections are small and the constraint algebra is closed. In this paper, we will focus on the latter case. Similar to the case with the holonomy corrections, we still adopt the slow-roll parameters defined in Eq.\ (\ref{epsilon}).

The inverse-volume corrections can be also introduced into equations governing the evolution of cosmological perturbations. In particular, it was found that, when inverse-volume corrections are present, the gauge-invariant comoving curvature perturbation $\mathcal{R}$ is conserved at large scales.  Such a feature of $\mathcal{R}$ strongly suggests that one can write a simple Mukhanov equation in the variable $\mu_k^{(s)}(\eta) \equiv z_s \mathcal{R}$, which is \cite{Bojowald2009c, Bojowald2011}
\bqn\lb{scalar-inv}
\frac{d^2\mu_k^{(s)}(\eta)}{d\eta^2}+\left(s^2(\eta) k^2-\frac{z_s(\eta)''}{z_s(\eta)}\right)\mu^{(s)}_k(\eta)=0,
\eqn
where
\bqn
z_s(\eta) \equiv \frac{a\dot \varphi}{H} \left[1+\frac{\alpha_0-2 \vartheta_0}{2} \delta_{\text{Pl}}\right],
\eqn
depends on the evolution of the background and
\bqn
s^2(\eta) \equiv 1+\chi \delta_{\text{Pl}},
\eqn
with
\bqn\lb{chi}
\chi \equiv \frac{\sigma \vartheta_0}{3} \left(\frac{\sigma}{6}+1\right)+\frac{\alpha_0}{2} \left(5-\frac{\sigma}{3}\right).
\eqn

For the cosmological tensor perturbations $h_k$, when the inverse-volume corrections are present, the corresponding Mukhanov equation for the variable $\mu^{(t)}_k(\eta) \equiv z_t h_k $ is written as \cite{Bojowald2011}
\bqn\lb{tensor-inv}
\frac{d^2\mu^{(t)}_k(\eta)}{d\eta^2}+\left(\alpha^2(\eta) k^2-\frac{z_t(\eta)''}{z_t(\eta)}\right)\mu^{(t)}_k(\eta)=0,
\eqn
with $\alpha(\eta)$ being given in Eq.\ (\ref{alpha}) and
\bqn
z_t(\eta) \equiv a \left(1-\frac{\alpha_0}{2} \delta_{\text{Pl}}\right).
\eqn

\subsection{Power spectra and spectral indices in the uniform asymptotic approximation}

Similar to the last section, to apply the uniform asymptotic approximation, we first write the equations of motion for both scalar (Eq.\ (\ref{scalar-inv})) and tensor perturbations (Eq.\ (\ref{tensor-inv})) into the standard form Eq.\ (\ref{eom}). Then, for the scalar perturbations, the functions $\hat g(y)$ and $q(y)$  must chosen as \cite{Zhu1,Zhu2},
\bqn
\lambda^2 \hat g(y)&=&-\frac{1}{k^2} \left(s^2(\eta) k^2-\frac{z_s''(\eta)}{z_s(\eta)}\right)+\frac{1}{4y^2},\\
q(y)&=& - \f{1}{4y^2},
\eqn
while for tensor perturbations one chooses
\bqn
\lambda^2 \hat g(y)&=&-\frac{1}{k^2} \left(\alpha^2(\eta) k^2-\frac{z_t''(\eta)}{z_t(\eta)}\right)+\frac{1}{4y^2},\\
q(y)&=& - \f{1}{4y^2}.
\eqn
In the slow-roll approximation, $z_s''/z_s$ and $z_t''/z_t$ can be casted in the form,
\bqn\lb{zsi}
\frac{z_s''(\eta)}{z_s(\eta)} &\simeq & a^2 H^2 \Big[2-\epsilon _1+\frac{3 \epsilon _2}{2}-\frac{\epsilon _1 \epsilon _2}{2}+\frac{\epsilon _2^2}{4}+\frac{\epsilon _2 \epsilon _3}{2}\nb\\
&&~~~~~~+f^{(s)}(\epsilon_i) \delta_{\text{Pl}}\Big],\nb\\
\eqn
and
\bqn\lb{zti}
\frac{z_t''(\eta)}{z_t(\eta)} & \simeq & a^2 H^2 \left[2-\epsilon_1 +f^{(t)}(\epsilon_i)\delta_{\text{Pl}}\right],
\eqn
where
\bqn
f^{(s)}(\epsilon_i)&\equiv&\frac{\sigma^2  (\sigma -3) \alpha _0}{4\epsilon_1} +\left( \frac{\sigma(\sigma -3) (\sigma +6)\vartheta_0}{12} +\frac{\sigma^2  \alpha _0}{4}\right)\nb\\
&&+\frac{\sigma(\sigma -3)\alpha _0}{4}  \frac{\epsilon _2}{\epsilon_1},\nb\\
f^{(t)}(\epsilon_i) &\equiv & \frac{\sigma (\sigma-3)}{2}\alpha_0.
\eqn
Because $\delta_{\text{Pl}} \sim y^{\sigma}$, it is convenient to write the function $\hat g(y)$ in a simplified form
\bqn\lb{gfunction}
\lambda^2 \hat g(y)=\frac{\nu^2(\eta)}{y^2}-1-\chi \delta_{\text{Pl}}+\frac{m(\eta)}{y^2} \delta_{\text{Pl}},
\eqn
where in a slow-roll background, $\nu(\eta)$ and $m(\eta)$ are slow-rolling variables depending on the types of the perturbations. In particular, for the scalar perturbation, $\chi$ is given by Eq.\ (\ref{chi}),  while for the tensor perturbation we shall replace $\chi$ by $2 \alpha_0$.

With the functions $\hat g(y)$ and $q(y)$ given in the above, we are in the position to calculate the power spectra and spectral indices from the general formulas Eq.\ (\ref{pw}) and Eq. (\ref{indices}). However, unlike the case for the holonomy corrections, in which $\nu(\eta)$ and $\Omega(\eta)$ are slow-rolling quantities, $\delta_{\text{Pl}}$ in Eq. (\ref{gfunction}) cannot be treated as a slow-rolling quantity during inflation. Consequently, the expansion of $ \sqrt{g(y)}=\sqrt{\lambda^2 \hat g(y)}$ in Eq.(\ref{ge}) cannot be directly applied to the function $g(y)$ of Eq.(\ref{gfunction}).

In order to apply for the formulas Eq.(\ref{pw}) and Eq.(\ref{indices}) to calculate inflationary observables with the inverse-volume corrections, let us first write $\delta_{\text{Pl}}$ as
\bqn
\delta_{\text{Pl}} &=&\left(\frac{a_{\text{Pl}}}{k}\right)^{\sigma} \left(\frac{H}{-a\eta H}\right)^\sigma y^{\sigma}=\epsilon_{\text{Pl}} \kappa(\eta) y^\sigma,
\eqn
with $\epsilon_{\text{Pl}}\equiv \left(\frac{a_{\text{Pl}}}{k}\right)^{\sigma} \ll 1$ and $\kappa(\eta) \equiv \left(\frac{H}{-a\eta H}\right)^\sigma$, and assume that {\em $\sigma$ is an integer within the range $0<\sigma \leq 6$}. With these conditions we can write the function $g(y)$ in the following form
\bqn
g(y)=\frac{y_0-y}{y^2} \left(h_0+h_1 y+\dots + h_\sigma y^\sigma +h_{\sigma+1} y^{\sigma+1}\right).\nb\\
\eqn
In the above $y_0=y_0(\eta)$ is assumed to be slow-rolling. Comparing the above form with Eq.\ (\ref{gfunction}) we find
\bqn
h_{\sigma+1}&=& \chi \epsilon_{\text{Pl}} \kappa,\nb\\
h_\sigma&=&\chi y_0 \epsilon_{\text{Pl}} \kappa,\nb\\
h_{\sigma-1}&=&  (\chi y_0-m y_0^{-1}) y_0 \epsilon_{\text{Pl}} \kappa,\nb\\
&&\dots\dots\nb\\
h_{i} &=& (\chi y_0-m y_0^{-1}) y_0^{\sigma-i} \epsilon_{\text{Pl}} \kappa,\;\;\;\;(\sigma>i\geq 2),\nb\\
&&\dots\dots\nb\\
h_2&=&  (\chi y_0-m y_0^{-1}) y_0^{\sigma-2}\epsilon_{\text{Pl}} \kappa,\nb\\
h_1&=& 1+ (\chi y_0-m y_0^{-1}) y_0^{\sigma-1}\epsilon_{\text{Pl}} \kappa,\nb\\
h_0&=& y_0+ (\chi y_0-m y_0^{-1}) y_0^{\sigma}\epsilon_{\text{Pl}} \kappa.
\eqn
Note that in the above we have assumed that $m=m(\eta)$ and $\kappa=\kappa(\eta)$ are slow-rolling quantities.  Then, expanding $\sqrt{g(y)}$ in terms of $\epsilon_{\text{Pl}}$ to the order $\mathcal{O}(\epsilon_{\text{Pl}})$, we find
\bqn
\sqrt{g(y)} &\simeq& \frac{ \sqrt{y_0-y}}{y} \Bigg\{\sqrt{y_0+y}\nb\\
&&+\frac{\epsilon_{\text{Pl}} \kappa}{2 \sqrt{y_0+y}} \Big[\chi (y_0^{\sigma+1}+y_0^\sigma y+\dots+y^{\sigma+1})\nb\\
&&~~~-\frac{m}{y_0^2} (y_0^{\sigma+1}+y_0^\sigma y+\cdots+y^{\sigma-1})\Big]\Bigg\}.\nb\\
\eqn
As we have assumed $0< \sigma \leq 6$, the two sequences in the above expressions are finite and can be expressed as
\bqn
&&y_0^{\sigma+1}+y_0^\sigma y+\dots+y^{\sigma+1}\nb\\
&&~~~~~~=y_0^{\sigma+1} \Big[1+\frac{y}{y_0}+a \left(\frac{y}{y_0}\right)^2+b\left(\frac{y}{y_0}\right)^3\nb\\
&&~~~~~~~~~~~~~~~~~+c\left(\frac{y}{y_0}\right)^4+d \left(\frac{y}{y_0}\right)^5\nb\\
&&~~~~~~~~~~~~~~~~~+e\left(\frac{y}{y_0}\right)^6+f \left(\frac{y}{y_0}\right)^7\Big],
\eqn
and
\bqn
&&y_0^{\sigma+1}+y_0^\sigma y+\dots+y_0^2y^{\sigma-1}\nb\\
&&~~~~~~=y_0^{\sigma+1} \Big[a+b \frac{y}{y_0}+c \left(\frac{y}{y_0}\right)^2+d\left(\frac{y}{y_0}\right)^3\nb\\
&&~~~~~~~~~~~~~~~~~+e\left(\frac{y}{y_0}\right)^4+f \left(\frac{y}{y_0}\right)^5\Big],
\eqn
where the relations between the values of $\sigma$ and $\{a,b,c,d,e,f\}$ is
\bqn
\sigma=6 \;\; &\Leftrightarrow& \;\;\{1,1,1,1,1,1\},\nb\\
\sigma=5 \;\; &\Leftrightarrow& \;\;\{1,1,1,1,1,0\},\nb\\
\sigma=4 \;\; &\Leftrightarrow& \;\;\{1,1,1,1,0,0\},\nb\\
\sigma=3 \;\; &\Leftrightarrow& \;\;\{1,1,1,0,0,0\},\nb\\
\sigma=2 \;\; &\Leftrightarrow& \;\;\{1,1,0,0,0,0\},\nb\\
\sigma=1 \;\; &\Leftrightarrow& \;\;\{1,0,0,0,0,0\}.
\eqn
Note that when $\sigma$ is an integer, one can always set $a=1$.

Now let us turn to the integral of $\sqrt{g(y)}$. In order to carry out the integration, we need also to specify the form of all the slow-rolling quantities $y_0(\eta)$, $m(\eta)$, and $\kappa(\eta)$. To do so, similar to the case with the holonomy corrections, it is convenient to expand all these quantities around the turning point $\bar y_0$ of $g(y)$ (i.e., $g(\bar y_0)=0$ with $\bar y_0(\eta_0)=-k \eta_0$) in the slow-roll inflation, i.e.,
\bqn\lb{slexpand}
y_0(\eta) &\simeq& \bar y_0+\bar y_1\ln\frac{y}{\bar y_0},\\
m(\eta) &\simeq& \bar m_0+\bar m_1\ln\frac{y}{\bar y_0},\\
\kappa(\eta) &\simeq& \bar \kappa_0+\bar \kappa_1\ln\frac{y}{\bar y_0}.
\eqn
In general the quantities $\bar y_0$, $\bar y_1$, $\bar m_0$, $\bar m_1$, $\bar \kappa_0$, $\bar \kappa_1$ are of order of
\bqn\lb{slexp}
\bar y_0 &\sim& \frac{3}{2}+\mathcal{O}(\epsilon_i)+\mathcal{O}\left(\frac{\epsilon_{\text{Pl}}}{\epsilon_i}\right),\nb\\
\bar y_1 &\sim & \mathcal{O}(\epsilon_i^2)+\mathcal{O}\left(\epsilon_{\text{Pl}}\right),\nb\\
\bar m_0 &\sim& \mathcal{O}(\frac{1}{\epsilon_i}),\;\;\;\bar m_1 \sim \mathcal{O}(1),\nb\\
\bar \kappa_0 &\sim& \bar H \mathcal{O}(1),\;\;\;\;\bar \kappa_1 \sim \bar H\mathcal{O}(\epsilon_i).
\eqn
As shown in \cite{Bojowald2011b}, when $\sigma$ is in the range of $1\leq \sigma \leq 6 $, the parameter of the inverse-volume corrections $H_\star^{\sigma}\epsilon_{\text{Pl}}$ is constrained by the observations that $H_\star^{\sigma}\epsilon_{\text{Pl}}$ should be $\lesssim 10^{-2}$. As we shall show below, in our calculations, the constraint is tighter because of $\epsilon_1^{-1}$ enhancement of the inverse-volume corrections. In this situation, as the slow-roll parameter is usually at order of $10^{-2}$, the correction term $H_\star^{\sigma}\epsilon_{\text{Pl}}$ is expected to be $\lesssim \epsilon_i^2$, which  usually is at the order of $10^{-4}$. Thus, in this paper we consider the inverse-volume corrections $H^\sigma \epsilon_{\text{Pl}}$ as the second-order in the slow-roll expansion.

Then, with these expansions we can perform the integral of  Eq.\ (\ref{pw}), and calculate the error control function $\mathscr{H}(+\infty)$. The results of the integral of $\sqrt{g(y)}$ and the error control function $\mathscr{H}(+\infty)$ are all presented in Appendix B. The slow-roll expansions of $\bar m_0$, $\bar m_1$, and $\bar \kappa_0,\;\bar \kappa_1$ etc are presented in Appendix C.

\subsection{Inflationary spectra for both scalar and tensor perturbations}

Let us first consider the scalar spectrum. With the slow-roll expansions of $\bar \nu_0,\;\bar\nu_1,\;\bar\nu_2,\;\bar\nu_3$, $\bar m_0,\;\bar m_1,\;\bar m_2$, etc, presented in Appendix C and the expression of $\bar y_0$ given in Eq.\ (\ref{bary0}), we find that up to the second-order in the slow-roll parameters, the scalar spectrum can be cast in the form
\bqn
\Delta_s^2(k) &\simeq&
\bar A_s
\Bigg\{
1-2 (1+\bar D_{\text{p}}) \bar \epsilon _1-\bar D_{\text{p}} \bar \epsilon _2\nb\\
&&~~~~+\left(2 \bar D_{\text{p}}+2 \bar D_{\text{p}}^2+\frac{\pi ^2}{2}-5+\bar \Delta _1\right) \bar \epsilon _1^2\nb\\
&&~~~~+\left(\bar D_{\text{p}}^2-\bar D_{\text{p}}+\frac{7 \pi ^2}{12}-8+\bar \Delta _1+2 \bar \Delta _2\right) \bar \epsilon _1 \bar \epsilon _2\nb\\
&&~~~~+\left(\frac{1}{2}\bar D_{\text{p}}^2+\frac{\pi ^2}{8}-\frac{3}{2}+\frac{1}{4}\Delta _1\right)\bar \epsilon _2^2\nb\\
&&~~~~+\left(\frac{\pi ^2}{24}+\bar\Delta _2-\frac{1}{2}\bar D_{\text{p}}^2\right) \bar \epsilon _2\bar  \epsilon _3\nb\\
&&~~~~+\epsilon_{\text{Pl}} \left(\frac{3\bar H}{2}\right)^\sigma\left(\frac{\mathcal{\bar Q}^{(s)}_{-1}}{\bar \epsilon_1}+\mathcal{\bar Q}^{(s)}_0+\frac{\mathcal{\bar Q}^{(s)}_{1} \bar \epsilon_2}{ \bar \epsilon_1}\right) \Bigg\}.\nb\\
\eqn
Then,  the corresponding scalar spectral index and running read
\bqn
n_s&\simeq&1-2 \bar \epsilon_1-\bar \epsilon _2-2 \bar \epsilon_1^2-\left(2\bar D_{\text{n}}+3\right) \bar \epsilon_1\bar \epsilon _2-\bar D_{\text{n}}\bar \epsilon _2 \bar \epsilon _3\nb\\
&&+\epsilon_{\text{Pl}} \left(\frac{3\bar H}{2} \right)^\sigma \left(\frac{\mathcal{\bar K}^{(s)}_{-1}}{\bar \epsilon_1}+\mathcal{\bar K}^{(s)}_0+\frac{\mathcal{\bar K}^{(s)}_1 \bar \epsilon_2}{ \bar \epsilon_1}\right),
\eqn
\bqn
\alpha_s &\simeq&
-2 \bar \epsilon _1 \bar \epsilon _2-\bar \epsilon _2 \bar \epsilon _3\nb\\
&&+\epsilon_{\text{Pl}} \left(\frac{3\bar H}{2}\right)^\sigma \left(\frac{\mathcal{\bar L}^{(s)}_{-1}}{\bar \epsilon_1}+\mathcal{\bar L}^{(s)}_0+\frac{\mathcal{\bar L}^{(s)}_1 \bar \epsilon_2}{\bar \epsilon_1}\right),
\eqn
where $\mathcal{\bar Q}_{-1}^{(s)},\;\mathcal{\bar Q}_{0}^{(s)},\;\mathcal{\bar Q}_{1}^{(s)}$, $\mathcal{\bar K}_{-1}^{(s)}, \mathcal{\bar K}_{0}^{(s)}, \mathcal{\bar K}_{1}^{(s)}$, and $\mathcal{\bar L}^{(s)}_{-1}, \mathcal{\bar L}_{0}^{(s)}, \mathcal{\bar L}_{1}^{(s)}$ are given in Table II.

Similar to the scalar perturbations, up to the second-order of the slow-roll parameters, the tensor spectrum, spectral index, and the running, are given, respectively, by
\bqn
\Delta_t^2(k) & \simeq &\bar A_t \Bigg\{1-2\left(1+\bar D_{\text{p}}\right) \bar \epsilon _1\nb\\
&&~~~~+\left(\bar \Delta _1+\frac{\pi ^2}{2}-5+2 \bar D_{\text{p}}+2 \bar D_{\text{p}}^2\right) \bar \epsilon _1^2\nb\\
&&~~~~+\left(2 \bar \Delta _2-2+\frac{\pi ^2}{12}-2 \bar D_{\text{p}}-\bar D_{\text{p}}^2\right)
   \bar \epsilon _1 \bar \epsilon _2\nb\\
&&~~~~+\epsilon_{\text{Pl}} \left(\frac{3}{2}\bar H\right)^\sigma \mathcal{\bar Q}^{(t)}_0\Bigg\},\\
n_t & \simeq & -2 \bar \epsilon _1-2 \bar \epsilon _1^2-2 \left(\bar D_{\text{n}}+1\right) \bar \epsilon _1 \bar \epsilon _2+\epsilon_{\text{Pl}} \left(\frac{3\bar H}{2}\right)^\sigma \mathcal{\bar K}_0^{(t)},\nb\\
\eqn
and
\bqn
\alpha_t &\simeq &
-2 \bar \epsilon _1 \bar \epsilon _2+\epsilon_{\text{Pl}} \left(\frac{3\bar H}{2}\right)^\sigma \mathcal{\bar L}^{(t)}_0,
\eqn
where $\mathcal{\bar Q}_{0}^{(t)}$, $\mathcal{\bar K}_{0}^{(t)}$, and $\mathcal{\bar L}^{(t)}_{0}$ are given in Table II.

\subsection{Evaluating at horizon crossing}

So far, we have obtained all the expressions of the power spectra, spectral indices, and running of spectral indices for both  scalar and tensor perturbations with the inverse-volume corrections. However, to compare with observations, we need to  express them  in terms of the slow-roll parameters which are evaluated at the time $\eta_\star$ when the scalar or tensor modes cross the horizon, i.e., $a(\eta_\star) H(\eta_\star) = s(\eta_\star) k$ for the scalar perturbations and  $a(\eta_\star) H(\eta_\star) = \alpha(\eta_\star) k$ for the tensor perturbations. Because in general the values of $s(\eta_\star)$ and $\alpha(\eta_\star)$ are different, for the scalar and tensor modes with the same wavenumber $k$, they cross the horizon at different times. When $s(\eta_\star) > \alpha(\eta_\star)$, the scalar mode leaves the horizon later than the tensor mode, and for $s(\eta_\star)<\alpha(\eta_\star)$, the scalar mode leaves the horizon earlier than the tensor. In this case, as we have pointed out in \cite{Uniform4}, caution must be taken for the evaluation time for all the inflationary observables. As we have two different horizon crossing times, it is reasonable to re-write all expressions in terms of quantities evaluated at the later time, i.e., we should evaluate all expressions at scalar-mode horizon crossing $a(\eta_\star) H(\eta_\star)= s(\eta_\star) k$ for $s(\eta_\star) > \alpha(\eta_\star)$ and at tensor-mode horizon crossing $a(\eta_\star) H(\eta_\star) = \alpha(\eta_\star) k$ for $s(\eta_\star) < \alpha(\eta_\star)$. However, detailed analysis shows that such a difference only contributes to the high-order terms in terms of the slow-roll parameters, which is beyond the approximation we consider here.
Thus, in this paper, we will not distinguish these two different cases and only consider the expansions at the time when the scalar mode crosses the Hubble horizon.

Then,  we shall re-write all the expressions in terms of quantities evaluated at the time when the scalar mode leaves the Hubble horizon $a(\eta_\star) H(\eta_\star)=s(\eta_\star) k$. Skipping all the tedious calculations, we find that the scalar spectrum can be written in the form
 \bqn\lb{sph}
 \Delta_s^2(k) &\simeq &A_s^{\star}\Bigg\{ 1-2\left(1+D_{\text{p}}^{\star}\right) \epsilon _{\text{$\star $1}}-D_{\text{p}}^{\star} \epsilon _{\text{$\star $2}}\nb\\
 &&+\left(2D_{\text{p}}^{\star 2}+2D_{\text{p}}^{\star}+\frac{\pi^2}{2}-5+\Delta_1^{\star}\right) \epsilon _{\text{$\star $1}}^2\nb\\
 &&+\left(\frac{D_{\text{p}}^{\star 2}}{2}+\frac{\pi^2}{8}-1+\frac{\Delta_1^{\star}}{4}\right) \epsilon _{\text{$\star $2}}^2\nb\\
&&+\left(D_{\text{p}}^{\star 2}-D_{\text{p}}^{\star}+\frac{7\pi ^2}{12}-7+\Delta_1^{\star}+2\Delta_2^{\star}\right) \epsilon _{\text{$\star $1}} \epsilon _{\text{$\star $2}}\nb\\
&&-\left(4 D_{\text{p}}^{\star}+6\right) \delta _{\text{$\star $H}} \epsilon _{\text{$\star $1}}+\left(\frac{\pi ^2}{24}-\frac{D_{\text{p}}^{\star 2}}{2}+\Delta_2^{\star}\right) \epsilon _{\text{$\star $2}} \epsilon _{\text{$\star $3}}\nb\\
 &&+\epsilon_{\text{Pl}} \left(\frac{3H_\star}{2}\right)^\sigma \Big[\frac{\mathcal{Q}_{-1}^{\star (s)}}{\epsilon_{\star 1}}+\mathcal{Q}_0^{\star (s)}+\frac{\mathcal{Q}_1^{\star (s)}\epsilon_{\star 2}}{\epsilon_{\star 1}}\Big]\Bigg\},\nb\\
 \eqn
where the subscript ``$\star$" denotes evaluation carried out at the horizon crossing, and
\bqn
\mathcal{Q}_{-1}^{\star (s)}&=& \mathcal{\bar Q}_{-1}^{(s)},\nb\\
\mathcal{Q}_{0}^{\star (s)} &=& \mathcal{\bar Q}^{(s)}_0+(\sigma +2) \mathcal{\bar Q}^{(s)}_{-1}\ln\frac{3}{2}+\frac{ \sigma ^2  (3-\sigma )\alpha _0}{9},\nb\\
\mathcal{Q}_{1}^{\star (s)} &=&2\mathcal{\bar Q}^{(s)}_{-1}\ln\frac{3}{2}+\mathcal{\bar Q}^{(s)}_1+\frac{\sigma ^2 (3-\sigma )\alpha _0}{18}.
\eqn
Now we turn to consider the scalar spectral index $n_s$, which can be expressed as
\bqn
n_s&\simeq&\lb{sih}
1-2 \epsilon_{\star 1}-\epsilon_{\star 2}-2 \epsilon_{\star 1}^2-(3+2 D^\star_{\text{n}}) \epsilon_{\star 1} \epsilon_{\star 2}-D^\star_{\text{n}} \epsilon_{\star 2} \epsilon_{\star 3}\nb\\
&&+\epsilon_{\text{Pl}} \left(\frac{3H_\star}{2}\right)^\sigma\Bigg\{\frac{\mathcal{K}_{-1}^{\star(s)}}{\epsilon _{\star 1}}+\mathcal{K}^{\star (s)}_0+\frac{\mathcal{K}^{\star(s)}_1 \epsilon_{\star 2}}{\epsilon_{\star 1}}\Bigg\},\nb\\
\eqn
where
\bqn
\mathcal{K}_{-1}^{\star (s)}&=&\mathcal{\bar K}_{-1}^{(s)},\nb\\
\mathcal{K}_{0}^{\star (s)}&=& \sigma \mathcal{\bar K}^{(s)}_{-1} \ln\frac{3}{2}+\mathcal{\bar K}^{(s)}_0,\nb\\
\mathcal{K}_{1}^{\star (s)}&=&\mathcal{\bar K}^{(s)}_1+\mathcal{\bar K}^{(s)}_{-1} \ln\frac{3}{2}.
\eqn
The running of the scalar spectral index reads
\bqn\lb{srh}
\alpha_s &\simeq&
-2 \epsilon_{\star1} \epsilon_{\star2}-\epsilon_{\star2} \epsilon_{\star3}\nb\\
&&+\epsilon_{\text{Pl}} \left(\frac{3H_\star }{2}\right)^\sigma\left\{\frac{\mathcal{L}_{-1}^{\star(s)}}{\epsilon _{\star 1}}+\mathcal{L}^{\star (s)}_0+\frac{\mathcal{L}^{\star(s)}_1 \epsilon_{\star 2}}{\epsilon_{\star 1}}\right\},\nb\\
\eqn
where
\bqn
\mathcal{L}_{-1}^{\star (s)}&=&\mathcal{\bar L}_{-1}^{(s)},\nb\\
\mathcal{L}_{0}^{\star (s)}&=& \sigma \mathcal{\bar L}^{(s)}_{-1} \ln\frac{3}{2}+\mathcal{\bar L}^{(s)}_0,\nb\\
\mathcal{L}_{1}^{\star (s)}&=&\mathcal{\bar L}^{(s)}_1+\mathcal{\bar L}^{(s)}_{-1} \ln\frac{3}{2}.
\eqn

For the tensor spectrum, we get
\bqn
\Delta_t^2(k) & \simeq &
A_t^\star \Bigg\{1-2\left(1+D_{\text{p}}^{\star}\right) \epsilon _{\text{$\star $1}}\nb\\
&&~~~~+\left(2D_{\text{p}}^{\star 2}+2D_{\text{p}}^{\star}+\frac{\pi^2}{2}-5+\Delta_1^{\star}\right) \epsilon _{\text{$\star $1}}^2\nb\\
&&~~~~+\left(-D_{\text{p}}^{\star 2}-2D_{\text{p}}^{\star}+\frac{\pi ^2}{12}-2+2\Delta_2^{\star}\right) \epsilon _{\text{$\star $1}} \epsilon _{\text{$\star $2}}\nb\\
&&~~~~+\epsilon_{\text{Pl}} \left(\frac{3 H_\star}{2}\right)^\sigma\mathcal{Q}_0^{\star (t)}\Bigg\},
\eqn
where $\mathcal{Q}_0^{\star (t)}=\mathcal{\bar Q}_0^{(t)}$. For the tensor spectral index, we find
\bqn
n_t &\simeq&
-2 \epsilon_{\star1}-2(1+D_{n}^\star) \epsilon_{\star1} \epsilon_{\star2}-2 \epsilon_{\star1}^2\nb\\
&&+\epsilon_{\text{Pl}} \left(\frac{3 H_\star}{2}\right)^\sigma \mathcal{K}_0^{\star (t)},
\eqn
where $\mathcal{K}_0^{\star (t)}=\mathcal{\bar K}_0^{ (t)}$. For the running of the tensor spectral index, we have
\bqn
\alpha_t &\simeq&
-2 \epsilon _{\star1} \epsilon _{\star2}+\epsilon_{\text{Pl}} \left(\frac{3 H_\star}{2}\right)^\sigma \mathcal{L}^{\star (t)}_0,
\eqn
where $\mathcal{L}_0^{\star (t)}=\mathcal{\bar L}_0^{ (t)}$.

Finally, with both scalar and tensor spectra given above, we can evaluate the tensor-to-scalar ratio at the horizon crossing $\eta_\star$, and we find that
\bqn
r &\simeq&
16 \epsilon _{\text{$\star $1}} \left\{1+D_\text{p}^{\star} \epsilon _{\star2}+\epsilon_{\text{Pl}} \left(\frac{3 H_\star}{2}\right)^\sigma \frac{\mathcal{Q}^{\star(s)}_{-1}}{\epsilon _{\star 1}}\right\}.\nb\\
\eqn

Some remarks about the spectra with the inverse-volume corrections now are in order. First, similar to the discussions given in \cite{Bojowald2011}, as $\epsilon_{\text{Pl}} \propto k^{-\sigma}$, both scalar and tensor spectra exhibit a deviation from the usual shape when $k$ is small enough, i.e., at large scales. Second, as a result of the above features in the spectra, the spectral indices, and especially the running of the spectral indices could be dominated by the quantum gravitational corrections at large scales. Such an interesting feature signals a qualitative departure from the inflation given in general relativity, and could be crucially important for further observational tests of constraints on quantum gravitational corrections.

Note that  in \cite{Bojowald2011,Bojowald2011b} the observables $n_s,\; n_t$ and  $r$ were calculated up to the first order of the slow-roll parameters.
Comparing their results with ours,  we find that  they are different. This is mainly due to  the following: (a) In \cite{Bojowald2011} the horizon crossing
was taken  as $k = {\cal{H}}$. However, due to the quantum gravitational effects, the dispersion relation is modified to the forms  (Eq.\ (\ref{scalar-inv})) and (Eq.\ (\ref{tensor-inv})), so the horizon crossing should be at $\omega_k =  {\cal{H}}$. (b) In \cite{Bojowald2011} the mode function was first obtained at two limits, ${k\gg {\cal{H}}}$ and ${k\ll {\cal{H}}}$, and then
 matched together at the horizon crossing  where $k \simeq  {\cal{H}}$.  This may lead  to  huge errors \cite{JM}, as neither $\mu_{k\gg {\cal{H}}}$ nor  $\mu_{k\ll {\cal{H}}}$   is a good approximation of the mode function $\mu_{k}$  at the horizon crossing.
 This is further supported  by considering the exact solution  for $\sigma=2$, to be presented in the next section.

\section{Inverse-volume correction for $\sigma=2$: exact solution}
\renewcommand{\theequation}{4.\arabic{equation}} \setcounter{equation}{0}

When $\sigma=2$, if we assume that all the slow-roll parameters are constants, the equations of motion (\ref{scalar-inv}) and (\ref{tensor-inv}) for both the scalar and tensor perturbations can be casted into the form
\bqn\lb{EoM1}
\mu_k''(\eta)+\left(\frac{a_0}{\eta^2}+a_1 k^2 +a_2 k^4 \eta^2\right)\mu_k(\eta)=0,
\eqn
where
\bqn
a_0\equiv \frac{1}{4}-\nu^2,\;\;
a_1 \equiv 1-m \epsilon_{\text{Pl}} \kappa,\;\;
a_2 \equiv \chi \epsilon_{\text{Pl}} \kappa.
\eqn
With $a_0,\;a_1$, and $a_2$ being constant, Eq. (\ref{EoM1}) can be solved exactly, whose solution reads
\bqn
\mu_k(\eta)&=& \frac{c_1}{\sqrt{-\eta}} WW\left(-\frac{ia_1}{4\sqrt{a_2}},\frac{\nu}{2},-i \sqrt{a_2} k^2 \eta^2\right)\nb\\
&&+\frac{c_2}{\sqrt{-\eta}} WM\left(\frac{ia_1}{4\sqrt{a_2}},\frac{\nu}{2},-i \sqrt{a_2} k^2 \eta^2\right),
\eqn
where $WW(b_1,b_2,z)$ and $WM(b_1,b_2,z)$ are the WhittakerW and WhittakerM functions, respectively.
To determine the coefficients $c_1$ and $c_2$, let us consider  the adiabatic initial condition,
\bqn\lb{ini}
\lim_{-k\eta \rightarrow +\infty}\mu_k(\eta)&=&\frac{1}{\sqrt{2 \omega_k(\eta)}} e^{-i \int \omega_k(\eta)d\eta}\nb\\
&
\simeq& \frac{1}{k a_2^{1/4}\sqrt{-2 \eta}} e^{i\frac{\sqrt{a_2} k^2 \eta^2}{2}}.
\eqn
The asymptotic forms of Whittaker functions $WW(b_1,b_2,z)$ and $WM(b_1,b_2,z)$ in the limit $|z| \rightarrow +\infty$, on the other hand,
are,
\bqn
WW(b_1,b_2,z) &\simeq& z^{b_1} e^{-\frac{z}{2}},\\
WM(b_1,b_2,z) &\simeq& \frac{\Gamma(1+2b_2)}{\Gamma(\frac{1}{2}+b_2+b_1)}z^{-b_1} e^{\frac{z}{2}}.
\eqn
Then, one arrives at
\bqn
\mu_k(\eta)&\simeq &\frac{c_1}{\sqrt{-\eta}} e^{\frac{i}{2} \sqrt{a_2}k^2 \eta^2} (-i \sqrt{a_2}k^2 \eta^2)^{\frac{ia_1}{4\sqrt{a_2}}}\nb\\
&&+\frac{c_2 \gamma}{\sqrt{-\eta}} e^{-\frac{i}{2} \sqrt{a_2}k^2 \eta^2} (-i \sqrt{a_2}k^2 \eta^2)^{\frac{ia_1}{4\sqrt{a_2}}},\nb\\
\eqn
where $\gamma\equiv \Gamma(1+\nu)/\Gamma(\frac{1+\nu}{2}+\frac{ia_1}{4\sqrt{a_2}})$. Comparing the above expressions with the initial condition and using the relation,
\bqn
(-i\sqrt{a_2}k^2\eta^2)^{\frac{ia_1}{4\sqrt{a_2}}}=e^{\frac{\pi a_1}{8\sqrt{a_2}}} e^{i\frac{a_1}{4\sqrt{a_2}} \ln(\sqrt{a_2}k^2\eta^2)},
\eqn
one gets $c_2=0$, and
\bqn
c_1=\frac{e^{-\frac{a_1 \pi}{8 \sqrt{a_2}}}}{\sqrt{2}k a_2^{1/4}}.
\eqn
One also can use the Wronskian condition
\bqn
\mu_k(\eta)\mu'^*_k(\eta)-\mu^{*}_k(\eta) \mu'_k(\eta)=i\nb
\eqn
to determine the coefficient $c_1$, which exactly gives the same result.
Thus,  with the initial condition (\ref{ini}) the solution reads
\bqn
\mu_k(\eta)= \frac{e^{-\frac{a_1 \pi}{8\sqrt{a_2}}}}{\sqrt{-2\eta}k a_2^{1/4}}WW\left(-\frac{ia_1}{4\sqrt{a_2}},\frac{\nu}{2},-i \sqrt{a_2} k^2 \eta^2\right).\nb\\
\eqn
Considering the asymptotic form in the limit $\eta\rightarrow 0^{-}$,
\bqn
&&WW\left(-\frac{ia_1}{4\sqrt{a_2}},\frac{\nu}{2},-i \sqrt{a_2} k^2 \eta^2\right)\nb\\
&&~~~~~~~~~~~~~~\simeq \frac{\Gamma(\nu)}{\Gamma(\frac{\nu+1}{2}+\frac{ia_1}{4\sqrt{a_2}})} (-i\sqrt{a_2}k^2\eta^2)^{\frac{1-\nu}{2}},\nb\\
\eqn
we find
\bqn
\mu_k(\eta)&\simeq& \frac{e^{-\frac{a_1 \pi}{8\sqrt{a_2}}}}{\sqrt{-2\eta}k a_2^{1/4}} \frac{\Gamma(\nu)}{\Gamma\left(\frac{\nu+1}{2}+\frac{i a_1}{4\sqrt{a_2}}\right)}\nb\\
&&\times \left(\sqrt{a_2} k^2 \eta^2 \right)^{\frac{1-\nu}{2}}.
\eqn
In the above we have ignored the irrelevant phase factor. At this position,  the power spectra can be computed in the limit $\eta\rightarrow 0^-$ as
\bqn
\Delta^2(k)\equiv \frac{k^3}{2\pi^2} \left|\frac{\mu_k(\eta)}{z(\eta)}\right|^2_{\eta\rightarrow 0^-}.
\eqn

For scalar perturbations, we find
\bqn
\Delta^2_s(k)&=&\frac{H^2}{16\pi^3 M_{\text{pl}}\epsilon_1}\Gamma^2(\nu) (a\eta H)^{-2} 2^{2\nu} (-k\eta)^{3-2\nu}a_1^{-\nu}\nb\\
&\simeq & \Delta_{\text{GR}}^2 \left(1+\nu m \kappa \epsilon_{\text{Pl}}-\f{2\nu(\nu^2-1)}{3}\chi \kappa \epsilon_{\text{Pl}}\right).\nb\\
\eqn
In the above, we have used the asymptotic formula of the Gamma function $\left|\Gamma\left(\frac{\nu+1}{2}+\frac{ia_1}{4\sqrt{a_2}}\right)\right|$. Then considering the slow-roll expansion of $\nu$, $m(\eta)$, and $\kappa$ for scalar perturbation,
\bqn
\nu_s&\simeq&\f{3}{2}+\epsilon_1+\frac{\epsilon_2}{2},\\
m_s&\simeq&\frac{\alpha_0}{\epsilon_1}+3\alpha_0-\frac{\alpha_0\epsilon_2}{2\epsilon_1},\\
\kappa&\simeq& H^{\sigma} (1-2\epsilon_1),
\eqn
one finds
\bqn
\Delta_s^2(k) \simeq \Delta_{\text{GR}}^2 \left[1+\alpha_0H^{2} \left(\frac{3}{2}\frac{1}{\epsilon_1}-\frac{5}{12}-\frac{1}{4} \frac{\epsilon_2}{\epsilon_1}\right)\epsilon_{\text{Pl}}\right].\nb\\
\eqn
It is easy to show that the coefficient of the leading order term ${\epsilon_{\text{Pl}}}/{\epsilon_1}$ is exactly consistent with the result obtained by the uniform asymptotic approximation, presented in the last section.

Let us turn to consider the tensor perturbations, for which we have
\bqn
\Delta_t^2(k)&\simeq& \frac{H^2}{8\pi^3 }\Gamma^2(\nu_t) (a\eta H)^{-1} 2^{2\nu} (-k\eta)^{3-2\nu} \nb\\
&&\times \left(1+\nu_t m_t \kappa \epsilon_{\text{Pl}}-\frac{4\nu_t (\nu_t^2-1)}{3} \alpha_0 \kappa \epsilon_{\text{Pl}}\right).\nb\\
\eqn
Considering the slow-roll expansions of $\nu_t(\eta)$, $m_t(\eta)$, and $\kappa$,
\bqn
\nu_t &\simeq& \frac{3}{2}+\epsilon_1,\\
m_t(\eta) &\simeq& \alpha_0+\alpha_0 \epsilon_1,
\eqn
we find
\bqn
\Delta_t^2(k)\simeq \Delta^2_{\text{GR}t}\left[1-\alpha_0 H^{2}\epsilon_{\text{Pl}}-3\alpha_0 H^2 \epsilon_1 \epsilon_{\text{Pl}}\right].
\eqn
One can check that the coefficient of the leading order term from the uniform asymptotic approximation obtained in the last section
 is $-\frac{183}{181} \sim - 1.011$, which is very close to the exact value $-1$, obtained from the above exact solution.
Therefore, the  results presented in the last section for $\sigma = 2$ are the same as these exact results obtained in this section within the errors allowed
 by our approximations.

\section{Detectability  of quantum gravitational effects}
\renewcommand{\theequation}{5.\arabic{equation}} \setcounter{equation}{0}

With the slow-roll conditions, the holonomy corrections are normally much weaker than the inverse-volume ones, and their effects in the early
universe are not expected to be observed in the near future \cite{QGEs}. However, this may not be the case for the  inverse-volume corrections \cite{Zhu3}. 
Therefore, in this section we shall consider only the latter. 
 
In general, with the spectral index and running given in Eqs.(\ref{sih}) and (\ref{srh}), the scalar spectrum can be expanded about a pivot scale $k_0$ as
\bqn
\ln{\Delta^2(k)} &\simeq& \ln{\Delta^2(k_0)}+\left[n_s(k_0)-1\right]\ln\left(\f{k}{k_0}\right)\nb\\&&+\frac{\alpha_s(k_0)}{2}\ln^2\left(\frac{k}{k_0}\right)+\sum_{n=3}^{\infty} \f{\alpha_s^{(n)}(k_0)}{n!} \ln^n\left(\f{k}{k_0}\right),\nb\\
\eqn
where up to the second-order approximations in terms of  the slow-roll parameters  and the leading order contribution from $\epsilon_{\text{Pl}}$ we have
\bqn
n_s-1& \simeq &-2 \epsilon _{\text{$\star $1}}-\epsilon _{\text{$\star $2}}-2 \epsilon _{\text{$\star $1}}^2-\left(3+2D^{\star}_{\text{n}} \right) \epsilon _{\text{$\star $1}} \epsilon _{\text{$\star $2}}\nb\\
&&-D^{\star}_{\text{n}}\epsilon _{\text{$\star $2}} \epsilon _{\text{$\star $3}}\nb+\epsilon_{\text{Pl}} \left(\frac{3H_\star}{2}\right)^\sigma \mathcal{K}_{-1}^{\star(s)}\epsilon_{\star1}^{-1},\nb\\
\alpha_s &\simeq& -2 \epsilon _{\text{$\star $1}} \epsilon _{\text{$\star $2}}-\epsilon _{\text{$\star $2}} \epsilon _{\text{$\star $3}}-\sigma \epsilon_{\text{Pl}} \left(\frac{3H_\star}{2}\right)^\sigma \mathcal{K}_{-1}^{\star(s)}\epsilon_{\star1}^{-1},\nb\\
\eqn
and
\bqn
\alpha^{(n)}_s(k) \simeq (-1)^{n-1}\sigma^{n-1} \epsilon_{\text{Pl}} \left(\frac{3H_\star}{2}\right)^\sigma \mathcal{K}_{-1}^{\star(s)}\epsilon_{\star1}^{-1}.
\eqn
Note that when $\sigma=3$, one has to replace $\mathcal{K}_{-1}^{\star (s)} \epsilon_{\star1}^{-1}$ by $\mathcal{K}_0^{\star (s)}$. Similar to
\cite{Bojowald2011b}, it is easy to find
\bqn
&&\sum_{m=3}^{\infty}\frac{\alpha^{(n)}_s(k_0)}{n!} \ln^n\frac{k}{k_0} \nb\\
&&~~~~~~~~=  -\epsilon_{\text{Pl}} \left(\frac{3H_\star}{2}\right)^\sigma \mathcal{K}_{-1}^{\star(s)}\epsilon_{\star1}^{-1} \nb\\
&&~~~~~~~~~~\times  \Big(\ln\f{k}{k_0}-\frac{\sigma}{2} \ln^2\f{k}{k_0}+\frac{e^{-\sigma \ln\f{k}{k_0}}-1}{\sigma} \Big).
\eqn

In order to carry out the CMB likelihood analysis, it is convenient to introduce the following potential slow-roll parameters,
\bqn
\epsilon_V \equiv \frac{M_{\text{Pl}}^2}{2} \frac{ V_\varphi^2}{V^2},\;\;\;\eta_V \equiv \frac{M_{\text{Pl}}^2 V_{\varphi \varphi}}{V}, \;\;\xi_V^2 \equiv \frac{M_{\text{Pl}}^4 V_{\varphi} V_{\varphi \varphi \varphi}}{V^2},\nb\\
\eqn
with which the scalar spectrum can be cast  in the form
\bqn
\Delta^2(k) &\simeq& \Delta^2(k_0) \exp\Bigg[\big(n_s(k_0)-1\big)\ln\f{k}{k_0} +\frac{\alpha_s(k_0)}{2}\ln^2\f{k}{k_0}\nb\\
&&\;\;\;\;~~+\frac{3^\sigma}{2^\sigma}\mathcal{K}_{-1}^{\star(s)} \frac{\epsilon_{\text{Pl}} H_\star^\sigma }{\epsilon_{V}}\nb\\
&&~~~~~~\times \Big(\ln\f{k}{k_0}-\frac{\sigma}{2} \ln^2\f{k}{k_0}+\frac{e^{-\sigma \ln\f{k}{k_0}}-1}{\sigma} \Big) \Bigg],\nb\\
\eqn
while the spectral index and its running can be written in the forms,
\bqn
n_s&\simeq& 1-6 \epsilon_V+2 \eta _V-\left(\frac{10}{3}+24 D^\star_{\text{n}}\right) \epsilon_V^2\nb\\
&&+(16 D^\star_{\text{n}}-2) \epsilon_V \eta_V+\frac{2 \eta_V^2}{3}+\left(\frac{2}{3}-2 D^\star_{\text{n}}\right) \xi_V^2\nb\\
&&+\frac{\epsilon_{\text{Pl}} H_\star^\sigma}{ \epsilon_{V}} \left\{\frac{3^\sigma}{2^\sigma}\mathcal{K}_{-1}^{\star(s)}+\frac{\sigma ^2(\sigma -3)\alpha_0 }{18}(3D^\star_{\text{n}}\sigma-\sigma -3)\right\},\nb\\
\eqn
and
\bqn
\alpha_s &\simeq & -2 \xi _V^2+16 \eta_V \epsilon_V-24 \epsilon _V^2\nb\\
&&-\frac{\sigma \epsilon_{\text{Pl}} H_\star^\sigma}{ \epsilon_{V}} \left\{\frac{3^\sigma}{2^\sigma}\mathcal{K}_{-1}^{\star(s)}-\frac{\sigma ^2(\sigma -3)\alpha_0 }{6}\right\}.
\eqn
Again remember that when $\sigma=3$ one has to replace $\mathcal{K}_{-1}^{\star (s)} \epsilon_{\star1}^{-1}$ by $\mathcal{K}_0^{\star (s)}$.

Let us  consider  the power-law potential
\bqn
\lb{PPs}
V(\varphi) =V_0 \varphi^{n},
\eqn
where $V_0$ and $n$ are constants. In this case, it follows that
\bqn
&&\epsilon_V =  \frac{M_{\text{Pl}}^2}{2} \frac{n^2}{\varphi^2}, \;\;\;\eta_V =M_{\text{Pl}}^2 \frac{n(n-1)}{\varphi^2},\nb\\
&&\xi_V^2 = M_{\text{Pl}}^4 \frac{n^2 (n-1)(n-2)}{\varphi^4},
\eqn
from which one can reduce the potential slow-roll parameters to one (i.e., $\epsilon_V$),
\bqn
\eta_V = \frac{2 (n-1)}{n} \epsilon_V, \;\; \xi_V^2 = \frac{4 (n-1)(n-2)}{n^2}\epsilon_V^2.
\eqn
It is also convenient to parameterize the inverse-volume corrections as
\bqn
\delta(k) = \alpha_0 \epsilon_{\text{Pl}} H_\star^\sigma .
\eqn
Thus, in the scalar power spectrum, with the power-law potential there are only two independent free parameters,
  $\epsilon_V(k_0)$ and $\delta(k_0)$.

Now let us turn to consider the observational effects of the inverse-volume corrections. When inverse-volume contributions vanish, we have $n_s=n_s(\epsilon_V)$ and $r=r(\epsilon_V)$. Thus it is easy to show that, up to the second-order of $\epsilon_V$, the relation
\cite{phi2},
\bqn
\lb{CNSTZ}
\Gamma_n (n_s, r) &\equiv& (n_s-1) + \frac{(2+n)r}{8n} \nb\\
&& + \frac{(3n^2 + 18n -4)(n_s-1)^2}{6(n+2)^2}   = 0, ~~~~
\eqn
holds precisely. The results from Planck 2015    are $n_s = 0.968 \pm 0.006$ and $r_{0.002} < 0.11 (95 \%$ CL) \cite{cosmo}, which yields $n_s \lesssim 1$.   In the forthcoming experiments, specially the   Stage
IV ones,  the errors  of the measurements on both $n_s$ and $r$ are  $ \sigma(n_s), \; \sigma (r) \le 10^{-3}$ \cite{S4-CMB}, which implies the error of the measurement of 
$\Gamma_n (n_s, r)$ is
\bq
\lb{Errors}
\sigma(\Gamma_n)   \le 10^{-3}.
\eq
Therefore, if any corrections to $n_s$ and $r$ lead to $\Gamma_n (n_s, r) \gtrsim 10^{-3}$, they should be within the range of detection of the current and forthcoming observations \cite{S4-CMB}. 

In particular,  when the inverse-volume corrections are taken into account ($\delta_{\text{Pl}} \not= 0$), we have $n_s = n_s(\epsilon_V, \epsilon_{\text{Pl}})$ and
$r = r(\epsilon_V,  \epsilon_{\text{Pl}})$, and  Eq.(\ref{CNSTZ})  is modified to,
\bq
\lb{CNSTZb}
\Gamma_n(n_s, r) = \mathcal{F}(\sigma) \frac{\delta(k)}{\epsilon_{V}},
\eq
where $\delta(k)\equiv \alpha_0 \epsilon_{\text{Pl}}H^{\sigma}$ and
\bqn
\lb{fs}
\mathcal{F}(\sigma) &=& \frac{3^\sigma}{2^\sigma}\mathcal{K}_{-1}^{\star(s)}+\frac{\sigma ^2(\sigma -3)\alpha_0 }{18}(3D^\star_{\text{n}}\sigma-\sigma -3).\nb\\
\eqn
Clearly, the right-hand side of the above equation represents the quantum gravitational effects from the inverse-volume corrections. If it is  equal or greater than $ {\cal{O}}(10^{-3})$, these effects shall be within the detection of the current or  forthcoming experiments. It is interesting to note that the quantum gravitational effects are enhanced by a factor of $\epsilon_{V}^{-1}$,
which  is absent in \cite{Bojowald2011}.

\begin{figure*}
\subfigure[$\;\;\;n=1$]{\label{zt1p1}
\includegraphics[width=6cm]{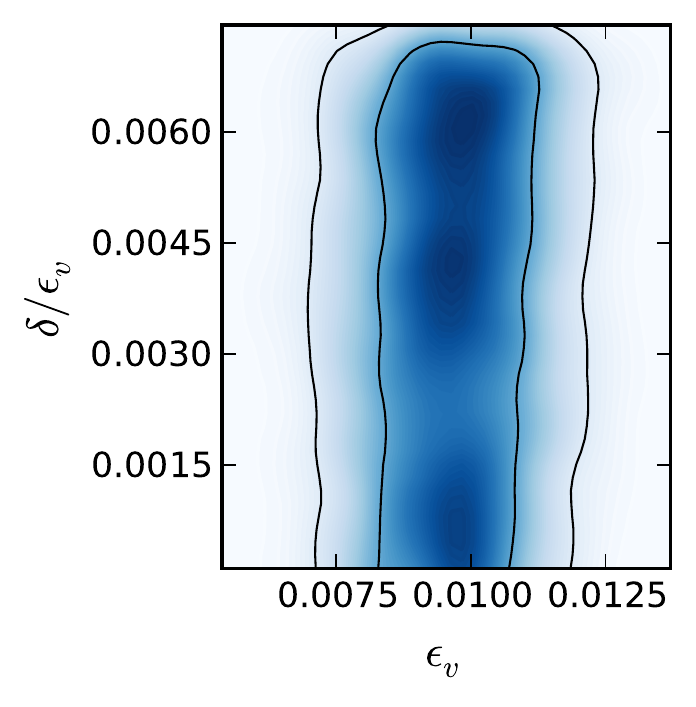}}
\subfigure[$\;\;\;n=\frac{2}{3}$]{\label{zt1p2d3}
\includegraphics[width=6cm]{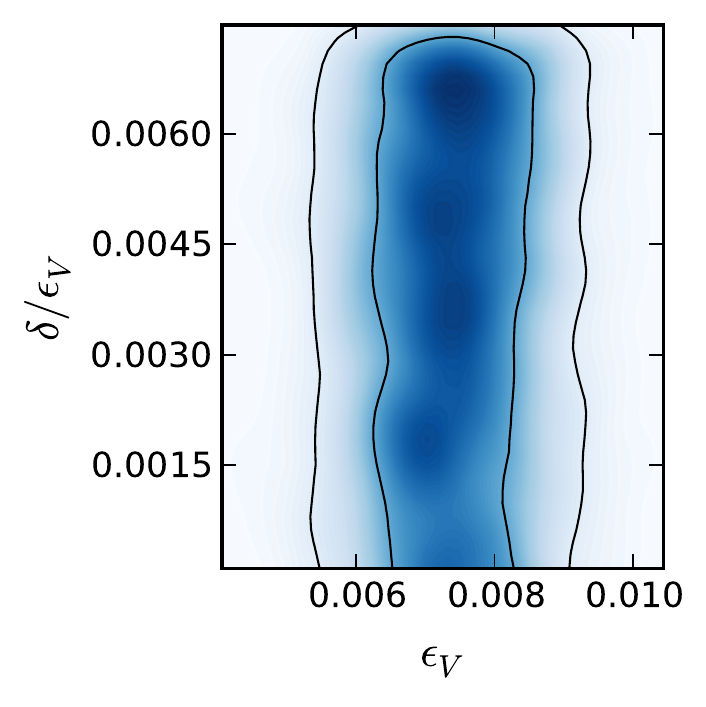}}\\
\subfigure[$\;\;\;n=\frac{3}{5}$]{\label{zt1p3d5}
\includegraphics[width=6cm]{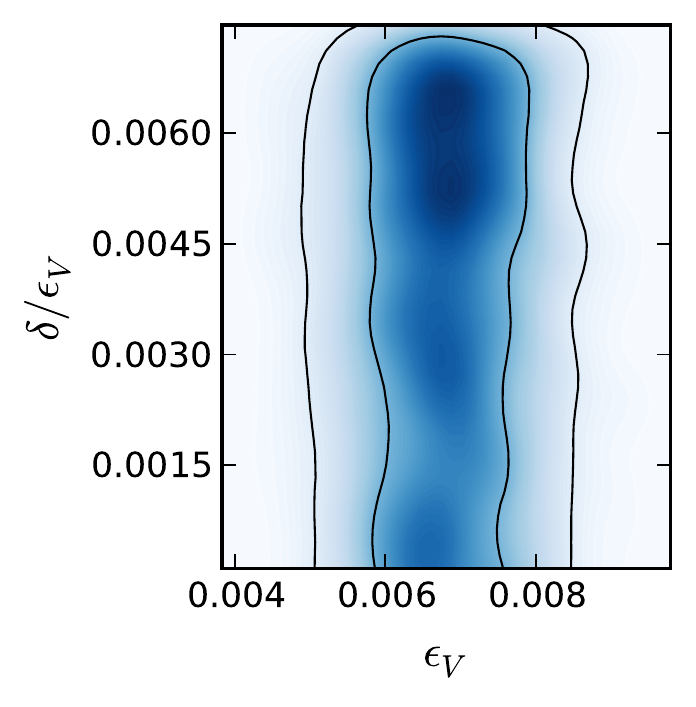}}
\subfigure[$\;\;\;n=\frac{1}{3}$]{\label{zt1[1d3}
\includegraphics[width=6cm]{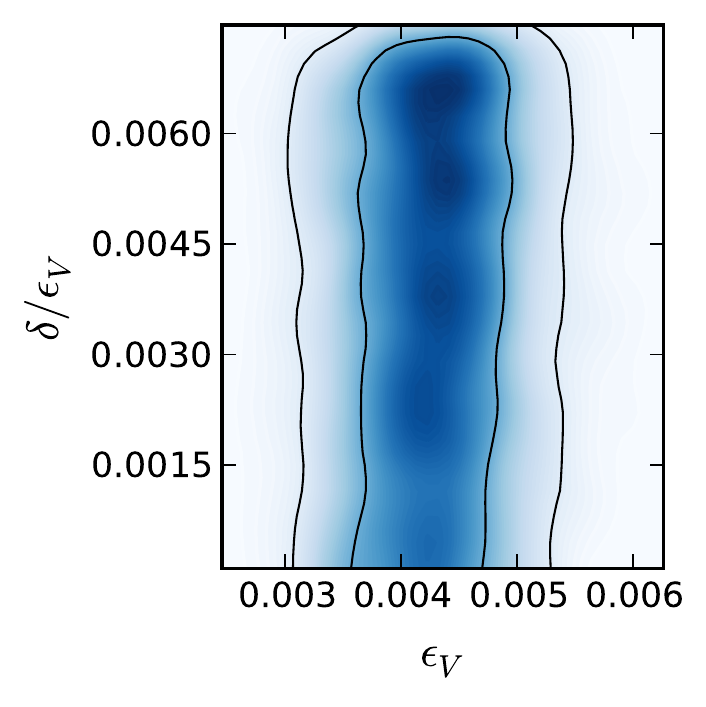}}
\caption{Two-dimensional marginalized distribution for the parameters $\delta/\epsilon_V$ and  $\epsilon_V$ at the pivot  $k_0=0.05\text{Mpc}^{-1}$ for the power-law potential with $n=1$, $n=\frac{2}{3}$, $n=\frac{3}{5}$, and $n=\frac{1}{3}$, respectively. The internal and external lines correspond to the confidence levels of  $68\%$ and $95\%$, respectively.} \label{fig1}
\end{figure*}

In the following, we run the Cosmological Monte Carlo (CosmoMC) code \cite{COSMOMC} with the Planck \cite{Planck2013}, BAO \cite{BAO2013}, and Supernova Legacy Survey \cite{SN} data for the power-law potential (\ref{PPs}) for
$n=1,\;\frac{3}{5},\;\frac{2}{3},\;\frac{1}{3}$, respectively. It is worthwhile to mention that all these potentials can be naturally realized in the axion monodromy inflation motivated by string/M theory \cite{axion}. In \cite{Zhu3}, by using CosmoMC code \cite{COSMOMC}, we already extracted constraints on loop quantum correction parameter $\delta_{\text{Pl}}/\epsilon_V$ and slow-roll parameter $\epsilon_V$ for $\sigma=1$ and $\sigma=2$ when $n=1$. It was noted that the constraint for $\sigma=2$ is much more tighter than the case of $\sigma=1$, and  makes the  quantum gravitational effects undetectable for models with $\sigma \ge 2$. It is interesting  to note that models
with small $\sigma$ are also favorable theoretically \cite{Bojowald2011,Bojowald2011b}. Therefore,   in the following we shall focus on the observational constraints for the case  $\sigma=1$.

We assume the flat cold dark matter model with effective number of neutrinos $N_{eff}=3.046$ and fix the total neutrino mass $\Sigma m_\nu=0.06 eV$. We vary the seven parameters: (i) baryon density parameter, $\Omega_bh^2$, (ii) dark matter density parameter, $\Omega_ch^2$, (iii) the ratio of the sound horiozn to the angular diameter, $\theta$, (iv) the reionization optical depth $\tau$, (v) $\delta(k_0)/\epsilon_{V}$, (vi) $\epsilon_{V}$, (vii) $\Delta_s^2(k_0)$. We take the pivot wave number $k_0 = 0.05 \; {\mbox{Mpc}}^{-1}$ used in Planck to constrain $\delta(k_0)$ and $\epsilon_V$. In Fig.\ref{fig1}, the constraints on ${\delta}/{\epsilon_{V}}$ and ${\epsilon_{V}}$  are given, respectively,  for $n=1$, $n=\frac{2}{3}$, $n=\frac{3}{5}$, and  $n=\frac{1}{3}$. In particular, we find that at $68\%$ C.L., 
\bqn
\delta(k_0) \lesssim 7.9\times 10^{-5} ,\;\;\;\;\;\text{with}\;\;n=1,\\
\delta(k_0) \lesssim 5.7\times 10^{-5} ,\;\;\;\;\;\text{with}\;\;n=\frac{2}{3},\\
\delta(k_0) \lesssim 5.4\times 10^{-5} ,\;\;\;\;\;\text{with}\;\;n=\frac{3}{5},\\
\delta(k_0) \lesssim 3.4\times 10^{-5} ,\;\;\;\;\;\text{with}\;\;n=\frac{1}{3},
\eqn
which are much tighter than those given in  \cite{Bojowald2011b}. In the above, to get the the bounds on $\delta(k_0)$, we have used the best-fit values of $\epsilon_V$, respectively for different $n$. It is also easy to see that the upper bound for both $\delta(k_0)$ and $\epsilon_V$ decrease slightly as $n$ decreases. However,  it is remarkable to note that the up bound on $\delta(k_0)/\epsilon_V$ is rather robust for different values of $n$, which is roughly
\bqn
\frac{\delta(k_0)}{\epsilon_V} \simeq {\cal{O}}(1) \times 10^{-3}\;\; (68\%\;\text{C.L.}).
\eqn
With such bounds, the gravitational quantum effects,  denoted by $\mathcal{F}(\sigma) \frac{\delta(k_0)}{\epsilon_V}$ in Eq. (\ref{CNSTZb}), could be within the range of the detection of the current and forthcoming cosmological experiments \cite{S4-CMB}. In addition, as we already pointed out in \cite{Zhu3}, the up bounds for $\frac{\delta(k_0)}{\epsilon_V}$ increase dramatically as $\sigma$ decreases. Thus it is very promising to expect the detectability of gravitational quantum effects for $\sigma \lesssim 1$  
 in the forthcoming cosmological experiments.

\section{Conclusions}
\renewcommand{\theequation}{5.\arabic{equation}} \setcounter{equation}{0}

The uniform asymptotic approximation method provides a powerful, systematically improvable, and error-controlled approach to construct accurate analytical solutions of linear perturbations. Its effectiveness has been verified by applying it to the inflation models with nonlinear dispersion relations \cite{Uniform3} and $k$-inflation \cite{Uniform4}. In this paper, we apply the high-order uniform asymptotic approximation to derive the inflationary observables for scalar and tensor perturbations in LQC with holonomy and inverse-volume quantum corrections. We obtain explicitly the analytical expressions of power spectra, spectral indices, and running of spectral indices up to the third-order approximation in terms of the parameters introduced in the uniform asymptotic approximation method. To this order, the upper error bounds are $\leq 0.15 \%$, accurate enough for the current and forthcoming experiments \cite{S4-CMB}. These expressions are all descibed in terms of the slow-roll parameters (up to the second-order) and  the parameters which represent the holonomy and inverse-volume quantum gravitational corrections.

For later applications of our results, we also rewrite all the inflationary observables including power spectra, spectral indices, and running of spectral indices for both scalar and tensor perturbations in terms of quantities evaluated at the time when the inflationary scalar (or tensor mode) crosses the Hubble horizon. With the resulting expressions, the tensor-to-scalar ratio is also obtained, and it is shown that the holonomy corrections do not contribute to the tensor-to-scalar ratio up to the second-order approximations. More interestingly, with the inverse-volume corrections, we find that both scalar and tensor spectra exhibit a deviation from the usual shape at large scales, which could be potentially important for the observational tests. As the uniform asymptotic approximate solution at the third-order has error bounds $\lesssim 0.15\%$, the inflationary observables obtained in the present paper represent the most accurate results obtained so far in the literature.

Utilizing the most accurate CMB, BAO and SN data  currently available publicly \cite{Planck2013,BAO2013,SN}, we also carry  out the CMB likelihood analysis, and find  the tightest constraints on $({\delta}(k_0), {\epsilon_{V}})$, obtained so far in the literature. Even with such tight constraints,  the quantum gravitational  effects due to the inverse-volume corrections of LQC can be within the range of the detection of the  current and forthcoming cosmological experiments \cite{S4-CMB}, provided that $\sigma \lesssim 1$.

\section*{Acknowledgements}

Part of the work was done when A.W. was visiting the State University of Rio de Janeiro (UERJ), and A.W. expresses his gratitude to UERJ and 
the colleagues there for their hospitality.  This work is supported in part by
Ci\^encia Sem Fronteiras, No. 004/2013 - DRI/CAPES, Brazil (A.W.);
NSFC No. 11375153 (A.W.), No. 11173021 (A.W.), No. 11047008 (T.Z.), No. 11105120 (T.Z.), and No. 11205133 (T.Z.), China.

\section*{Appendix A: The uniform asymptotic approximation}
\renewcommand{\theequation}{A.\arabic{equation}} \setcounter{equation}{0}

In this section, we present a brief introduction of the {\em uniform asymptotic approximation method} and its applications to the inflationary cosmology for the cases where the dispersion relation has only a single turning point. For details, we refer readers to our original papers \cite{Zhu1, Zhu2,Uniform3, Uniform4}.

\subsection*{A. The approximate solution of the mode function}

In the uniform asymptotic approximation method \cite{uniformPRL,Olver1974,Zhu1,Zhu2}, one usually works with the second-order differential equation,
\bqn\lb{eom}
\frac{d^2\mu_k(y)}{dy^2}=\left[\lambda^2\hat g(y)+q(y)\right]\mu_k(y).
\eqn
In the above the parameter $\lambda$ is used to trace the order of the uniform approximations. Usually $\lambda$ is supposed to be large, and it also can be absorbed into $\hat g(y)$. Thus when we turn to determine the final results, we can set $\lambda=1$ for the sake of simplicity. For convenience, we also use the notation $g(y)=\lambda^2 \hat g(y)$. Specific to the cosmological applications, $\mu_k(y)$ represents the inflationary mode function for cosmological scalar or tensor perturbations, and one can identify
\bqn
\lambda^2\hat g(y)+q(y) \equiv -\frac{1}{k^2} \left(\omega^2_k(\eta)-\frac{z''(\eta)}{z(\eta)}\right),
\eqn
where $\omega_k^2(\eta)$ is the associated dispersion relation for the inflationary mode function $\mu_k(y)$ and $z(\eta)$ depends on the cosmological background evolution. In most of the cases, $\hat g(y)$ and $q(y)$ have two poles (singularities): one is at $y=0^+$ and the other is at $y=+\infty$. As we discussed in \cite{Zhu2} (see also \cite{uniformPRL,Olver1974}), if these two poles are both second-order or higher, one needs to choose
\bq
\lb{qF}
q(y)=- \frac{1}{4y^2},
\eq
for ensuring the convergence of the error control functions. In this paper we shall restrict our investigations to this case. In addition, the function $\hat g(y)$ can vanish at various points, which are called turning points or zeros, and the approximate solution of the mode function $\mu_k(y)$ depends on the behavior of $\hat g(y)$ and $q(y)$ near these turning points.

To proceed further, let us first introduce the Liouville transformation with two new variables $U(\xi)$ and $\xi$ via the relations,
\bqn
\lb{Olver trans}
U(\xi)&=& \chi^{1/4} \mu_k(y),\;\;\; \xi'^2 =  \frac{|\hat g(y)|}{f^{(1)}(\xi)^2},
\eqn
where $ \chi \equiv \xi'^2,\; \xi'=d\xi/dy$, and
\bqn
\lb{OlverTransB}
f(\xi)&=& \int^y \sqrt{|\hat g(y)|} dy,\;\;\;  f^{(1)}(\xi)=\frac{df(\xi)}{d\xi}.
\eqn
Note that $\chi$ must be regular and not vanish in the intervals of interest. Consequently, $f(\xi)$ must be chosen so that
$f^{(1)}(\xi)$ has zeros and singularities of the same type as that of $\hat g(y)$. As shown below, such a requirement plays an essential role in determining the approximate solutions. In terms of $U$ and $\xi$, Eq.\ (\ref{eom}) takes the form
\bqn\lb{eomU}
\frac{d^2 U}{d\xi^2}&=&\left[\pm \lambda^2 f^{(1)}(\xi)^2+\psi(\xi)\right]U,
\eqn
where
\bqn\lb{psi}
\psi(\xi)=\frac{q(y)}{\chi}-\chi^{-3/4} \frac{d^2(\chi^{-1/4})}{dy^2},
\eqn
and the signs ``$\pm$" correspond to $\hat g(y)>0$ and $\hat g(y)<0$, respectively. Considering $\psi(\xi) =0$ as the first-order approximation,
one can choose $f^{(1)}(\xi)$ so that the first-order approximation can be as close
to the exact solution as possible with the guidelines of the error functions constructed below, and then  solve it in terms of known functions.
Clearly, such a choice  sensitively  depends on the behavior of the functions $\hat g(y)$ and $q(y)$ near  the poles and turning points.

In  this paper, we consider only the case in which $\hat g(y)$ has only one single turning point $\bar{y}_0$
(for $\hat g(y)$ having several different turning points or one multiple-turning point,
see \cite{Zhu2}), i.e., $\hat g(\bar y_0)=0$.
In  this case we can choose
\bqn
f^{(1)}(\xi)^2=\pm \xi,
\eqn
where $\xi=\xi(y)$ is a monotone decreasing function, and $\pm$ correspond to $\hat g(y)\geq 0$ and $\hat g(y) \leq 0$, respectively.
Following Olver  \cite{Olver1974}, the general solution of Eq.\ (\ref{eomU}) can be written as
\bqn\lb{appro}
U(\xi)&=&\alpha_0 \Bigg[\text{Ai}(\lambda^{2/3} \xi) \sum_{s=0}^{n} \frac{A_s(\xi)}{\lambda^{2s}}\nb\\
&&~~~~~~~+\frac{\text{Ai}'(\lambda^{2/3}\xi)}{\lambda^{4/3}} \sum_{s=0}^{n-1} \frac{B_s(\xi)}{\lambda^{2s}}+\epsilon_3^{(2n+1)}\Bigg]\nb\\
&&+\beta_0 \Bigg[\text{Bi}(\lambda^{2/3} \xi) \sum_{s=0}^{n} \frac{A_s(\xi)}{\lambda^{2s}}\nb\\
&&~~~~~~~~+\frac{\text{Bi}'(\lambda^{2/3}\xi)}{\lambda^{4/3}} \sum_{s=0}^{n-1} \frac{B_s(\xi)}{\lambda^{2s}}+\epsilon_4^{(2n+1)}\Bigg],\nb\\
\eqn
where $\text{Ai}(x)$ and $\text{Bi}(x)$ represent the Airy  functions, $\epsilon_3^{(2n+1)}$ and $\epsilon_{4}^{(2n+1)}$ are errors of the approximate solution, and
\bqn\lb{AB}
&& A_0(\xi)=1,\;\;\nb\\
&&B_s(\xi)=\frac{\pm 1}{2 (\pm \xi)^{1/2}}\int_0^\xi \{\psi(v) A_s(v)-A''_s(v)\}\frac{dv}{(\pm v)^{1/2}},\nb\\
&&A_{s+1}(\xi)=-\frac{1}{2} B'_s(\xi)+\frac{1}{2} \int \psi(v) B_s(v) dv,\nb\\
\eqn
where $\pm$ correspond to $\xi \geq 0$ and $\xi\leq 0$, respectively. The error bounds of $\epsilon_3^{(2n+1)}$ and $\epsilon_4^{(2n+1)}$ can be expressed as
\bqn\lb{error}
&&\frac{\epsilon_{3}^{(2n+1)}}{M(\lambda^{2/3} \xi)}, \;\;\frac{\partial \epsilon_{3}^{(2n+1)}/\partial\xi}{\lambda^{2/3} N(\lambda^{2/3}\xi)}\nb\\
&& ~~~~~\leq 2 E^{-1}(\lambda^{2/3}\xi)  \exp{\left[\frac{2\kappa_0 \mathscr{V}_{\alpha,\xi}(|\xi^{1/2}|B_0)}{\lambda}\right]} \nb\\
&&~~~~~~~~~~~\times \frac{\mathscr{V}_{\alpha,\xi}(|\xi^{1/2}|B_n)}{\lambda^{2n+1}},\nb\\
&&\frac{\epsilon_{4}^{(2n+1)}}{M(\lambda^{2/3} \xi)}, \;\;\frac{\partial \epsilon_{4}^{(2n+1)}/\partial\xi}{\lambda^{2/3} N(\lambda^{2/3}\xi)} \nb\\
&&~~~~~\leq 2 E(\lambda^{2/3}\xi) \exp{\left[\frac{2\kappa_0 \mathscr{V}_{\xi,\beta}(|\xi^{1/2}|B_0)}{\lambda}\right]} \nb\\
&&~~~~~~~~~~~\times\frac{\mathscr{V}_{\xi,\beta}(|\xi^{1/2}|B_n)}{\lambda^{2n+1}},
\eqn
where the definitions of $M(x)$, $N(x)$, $\kappa_0$, and $\mathscr{V}_{a,b}(x)$ can be found in \cite{Zhu2}.

\subsection*{B. Power spectra and spectral indices up to the third-order}

With the approximate solution given in the last section, now let us begin to calculate the inflationary power spectra and spectral indices from the approximate solution.
We assume that the universe was initially at the adiabatic vacuum,
\bqn
\lim_{y\to +\infty} \mu_k(y)=\lim_{y\to +\infty} \frac{1}{\sqrt{2 \omega_k(\eta)}} e^{-i \int \omega_k(\eta) d\eta}.
\eqn
Then,  we need to match this initial state with the approximate solution (\ref{appro}).  However, the approximate solution (\ref{appro}) involves many high-order terms, which are  complicated
and not easy to handle. In order to simplify them, we  first study their behavior in the limit $y\rightarrow +\infty$. Let us  start with  the $B_0(\xi)$ term in Eq.\ (\ref{AB}), which satisfies
\bqn
B_0(\xi)=-\frac{1}{2\sqrt{-\xi}} \int_{0}^{\xi} \frac{\psi(v)}{\sqrt{-v}}dv=-\frac{\mathscr{H}(\xi)}{2\sqrt{-\xi}},
\eqn
where $\mathscr{H}(\xi)\equiv \int_{0}^{\xi} dv \psi(v)/|v|^{1/2}$ is the associated error control function of the approximate solution (\ref{appro}), and
in the above we have used $A_0(\xi)=1$. The error control function $\mathscr{H}(\xi)$ is well behaved around the turning point $\bar y_0$ and converges  when $y\to +\infty$. As a result, we have
\bqn
\lim_{y\rightarrow +\infty} B_0(\xi) =-\frac{\mathscr{H}(-\infty)}{2\sqrt{-\xi}}.
\eqn
Then, let us  turn to  $A_1$, which is
\bqn
A_1(\xi)=-\frac{1}{2} B_0'(\xi)+\frac{1}{2}\int_{0}^{\xi} \psi(v)B_0(v)dv.
\eqn
In the limit $y\rightarrow +\infty$,  $B_0'(\xi)$ vanishes, and we find
\bqn
\lim_{y\to+\infty} A_1(\xi) &=&-\frac{1}{2} \int_{0}^{\xi} \frac{\psi(v)}{\sqrt{-v}} \left[\frac{1}{2}\int_{0}^{v} \frac{\psi(u)}{\sqrt{-u}} du\right] dv\nb\\
&=&-\frac{1}{2} \left[\frac{\mathscr{H}(-\infty)}{2}\right]^2.
\eqn
Note that in the above we have used the formula
\bqn
&&n! \int_{\xi_0}^{\xi} f(\xi_n) \int_{\xi_0}^{\xi_n} f(\xi_{n-1})\cdots \int_{\xi_0}^{\xi_2}f(\xi_1)d\xi_1d\xi_2\cdots d\xi_n\nb\\
&&~~~~~~~~~~~~~~~~~~~~~~~~~~~~~~=\left[\int_{\xi_0}^{\xi} f(v)dv\right]^n.
\eqn
Thus,  up to the third-order, we have
\bqn
A_0(\xi)+\frac{A_1(\xi)}{\lambda^2} &=&1-\frac{1}{2\lambda^2} \left[\frac{\mathscr{H}(-\infty)}{2}\right]^2+\mathcal{O}\left(\frac{1}{\lambda^3}\right),\;\;\;\;\;\nb\\
\frac{B_0(\xi)}{\lambda}&=& -\frac{1}{\sqrt{-\xi}} \frac{\mathscr{H}(-\infty)}{2\lambda}+\mathcal{O}\left(\frac{1}{\lambda^3}\right).
\eqn

Using the asymptotic form of Airy  functions in the limit $\xi\to -\infty$, and comparing the solution $\mu_k(y)$ with the initial state, we obtain
\bqn
\alpha_0=\sqrt{\frac{\pi}{2k}} \frac{\lambda^{1/3}}{(A_0+A_1/\lambda^2)-i \sqrt{-\xi} B_0/\lambda},\nb\\
\beta_0=i \sqrt{\frac{\pi}{2k}} \frac{\lambda^{1/3}}{(A_0+A_1/\lambda^2)-i \sqrt{-\xi} B_0/\lambda},
\eqn
where we have
\bqn
(A_0+A_1/\lambda^2)-i \sqrt{-\xi} B_0/\lambda =(1+\mathcal{O}(1/\lambda^3)) e^{i\theta}.  ~~~~~~~
\eqn
Here $\theta$ is an irrelevant phase factor, and without loss of generality, we can set $\theta=0$. Thus, we  finally get
\bqn
\frac{\alpha_0}{\lambda^{1/3}}=\sqrt{\frac{\pi}{2k}},\;\;\;\frac{\beta_0}{\lambda^{1/3}}=i\sqrt{\frac{\pi}{2k}}.
\eqn

After determining the coefficients $\alpha_0$ and $\beta_0$, we can calculate the power spectra of the perturbations. As $y \rightarrow 0$, only  the growing mode is relevant. Thus we have
\bqn
\mu_k(y)&\simeq& \beta_0 \left(\frac{\xi}{\hat{g}(y)}\right)^{1/4}\Bigg[\text{Bi}(\lambda^{2/3}\xi)\sum_{s=0}^{+\infty} \frac{B_s(\xi)}{\lambda^{2s}}\nb\\
&&\;\;\;\;\;\;\;\;\;\;\;\;+\frac{\lambda^{2/3}\text{Bi}'(\lambda^{2/3}\xi)}{\lambda^2}\sum_{s=0}^{+\infty}\frac{B_s(\xi)}{\lambda^{2s}}\Bigg].\;\;\;\;\;\;\;\;
\eqn
In order to calculate  the power spectra to higher order, let us first consider the $B_0(\xi)$ term, which satisfies
\bqn
\lim_{y\to 0 }B_0(\xi)=\frac{1}{2\xi^{1/2}} \int_{0}^{\xi} \frac{\psi(v)}{v^{1/2}}dv=\frac{\mathscr{H}(+\infty)}{2\xi^{1/2}}.
\eqn
In the above we had used the relation $\xi^{1/2}d\xi=-\sqrt{\hat{g}}dy$. Knowing the $B_0$ term, we can get the $A_1$ term, which is
\bqn
\lim_{y\to 0} A_1(\xi)&=&\frac{1}{4} \int_{0}^{\xi} \frac{\psi(v)}{v^{1/2}}  \int_{0}^{v} \frac{\psi(u)}{u^{1/2}} du dv\nb\\
&=&\frac{1}{2} \left[\frac{\mathscr{H}(+\infty)}{2}\right]^2.
\eqn

Thus up to the third order and considering the asymptotic forms of the Airy functions in the limit $\xi\rightarrow +\infty$, we find
\bqn
\lim_{y\rightarrow 0} \mu_k(y)&=& \frac{\beta_0 e^{\frac{2}{3}\lambda \xi^{2/3} } }{ \lambda^{1/6} \hat{g}^{1/4} \pi^{1/2} }
\Bigg[1+\frac{\mathscr{H}(+\infty)}{2\lambda}+\frac{\mathscr{H}(+\infty)^2}{8\lambda^2}\nb\\
&&~~~~~~~~~~~~~~~~~~~~~~~~+\mathcal{O}(1/\lambda^3)\Bigg].
\eqn
Then,  the power spectra can be calculated, and is given by
\bqn\lb{pw}
\Delta^2(k)&\equiv& \frac{k^3}{2\pi^2} \left|\frac{\mu_k(y)}{z}\right|^2_{y\to 0^{+}}\nb\\
&\simeq&\frac{k^2}{4\pi^2}\frac{-k\eta}{z^2(\eta) \nu(\eta)}\exp\left(2 \int_y^{\bar y_0}\sqrt{{g}(\hat{y})}d\hat{y}\right)\nb\\
&&\;\times \left[1+\frac{\mathscr{H}(+\infty)}{\lambda}+\frac{\mathscr{H}^2(+\infty)}{2\lambda^2}+\mathcal{O}(1/\lambda^3)\right].\nb\\
\eqn
From the power spectra presented above, one can get the general expression of the spectral indices, which now is given by
\bqn\lb{indices}
n-1&\equiv&\frac{d\ln \Delta^2(k)}{d\ln k}\nb\\
&\simeq&3+2 \frac{d}{d\ln k} \int_y^{\bar y_0} \sqrt{g(\hat y)} d\hat y+\frac{1}{\lambda}\frac{d\mathscr{H}(+\infty)}{d\ln k}\nb\\
&&+\mathcal{O}\left(\frac{1}{\lambda^3}\right),
\eqn
where the last term in the above expression represents the second- and third-order approximations. It should be noted that the above results represent the most general expressions of the power spectra and  spectral indices of perturbations for the case that has only one-turning point.

\section*{Appendix B: Integral of $\sqrt{g(y)}$ and the error control function $\mathscr{H}(+\infty)$ with inverse-volume corrections}
\renewcommand{\theequation}{B.\arabic{equation}} \setcounter{equation}{0}

In general, the integral of $\sqrt{g}$ can be divided into two parts,
\bqn
\int_y^{\bar y_0} \sqrt{g(y)}dy \simeq I_1+I_2,
\eqn
where after some tedious calculations we find
\bqn\lb{int-g-inv}
\lim_{y\to 0} I_1 &=&\bar y_0 \Big[-1-\ln\frac{y}{2\bar y_0}+\left(-\frac{\pi }{2}+\ln2-\ln\frac{y}{\bar y_0}\right) A_0\nb\\
&&~~~~+\frac{1}{2} (-2+\pi )A_1+\left(1-\frac{\pi }{4}\right) A_2\nb\\
&&~~~~+\left(-\frac{2}{3}+\frac{\pi }{4}\right)A_3+\left(\frac{2}{3}-\frac{3 \pi }{16}\right) A_4\nb\\
&&~~~~+\left(-\frac{8}{15}+\frac{3 \pi}{16}\right) A_5+\left(\frac{8}{15}-\frac{5 \pi }{32}\right)A_6\nb\\
&&~~~~+\left(-\frac{16}{35}+\frac{5 \pi }{32}\right) A_7\Big],
\eqn
\bqn
\lim_{y\to 0} I_2 &=& \left(-\frac{\pi ^2}{24}+\frac{\ln^22}{2}-\frac{1}{2}\ln^2\frac{y}{\bar y_0}\right)\bar y_1\nb\\
&&+\bar y_0\Bigg[\left(-\frac{\pi ^2}{24}-\ln2+\frac{\ln^22}{2}-\frac{1}{2}\ln^2\frac{y}{\bar y_0}\right)B_0\nb\\
&&-\frac{\pi  }{2}B_1-B_2\ln2 +\left(-\frac{\pi }{2}+\frac{\pi  \ln2}{2} \right) B_3\nb\\
&&+(1-\ln4) B_4+\left(-\frac{5 \pi }{8}+\frac{3\pi  \ln2}{4} \right) B_5\nb\\
&&+\left(\frac{14}{9}-\frac{8 \ln2}{3}\right) B_6+\left(\frac{15\pi  \ln2}{16}-\frac{47 \pi}{64}\right) B_7\nb\\
&&+\left(\frac{148}{75}-\frac{16 \ln2}{5}\right) B_8
\Bigg],
\eqn
where $A_0,\cdots,A_7$, $B_0, \cdots,B_{8}$, and $C_0$ are given by
\bqn
A_0&=&\frac{1}{2} \bar y_0^{-2+\sigma } \left(-a \bar m_0+\chi  \bar y_0^2\right)  \epsilon _{\text{Pl}} \bar \kappa _0,\nb\\
A_1&=& \frac{1}{2} \bar y_0^{-2+\sigma } \left(-b \bar m_0+\chi  \bar y_0^2\right) \epsilon _{\text{Pl}}\bar \kappa _0,\nb\\
A_2 &=& \frac{1}{2} y_0^{-2+\sigma } \left(-c \bar m_0+a \chi  \bar y_0^2\right) \epsilon _{\text{Pl}} \bar \kappa _0,\nb\\
A_3&=&\frac{1}{2} \bar \kappa _0 \epsilon _{\text{Pl}} y_0^{\sigma -2} \left(b \chi  y_0^2-d \bar m_0\right),\nb\\
A_4&=& \frac{1}{2} \bar \kappa _0  \epsilon _{\text{Pl}} \bar y_0^{\sigma -2} \left(c \chi  \bar y_0^2-e \bar m_0\right),\nb\\
A_5&=& \frac{1}{2} \bar \kappa _0  \epsilon _{\text{Pl}} \bar y_0^{\sigma -2} \left(d \chi  \bar y_0^2-f \bar m_0\right),\nb\\
A_6&=& \frac{1}{2} e \bar \kappa _0 \chi \epsilon _{\text{Pl}} \bar y_0^{\sigma },\nb\\
A_7&=&\frac{1}{2} f \bar \kappa _0 \chi  \epsilon _{\text{Pl}} \bar y_0^{\sigma },
\eqn
and
\bqn
B_0&=&-\frac{a}{2} \epsilon _{\text{Pl}} \bar y_0^{\sigma -2} \left(\bar \kappa _0 \bar m_1+\bar \kappa _1 \bar m_0\right),\nb\\
B_1&=&\frac{1}{2} (a-b) \epsilon _{\text{Pl}} \bar y_0^{\sigma -2} \left(\bar \kappa _0 \bar m_1+\bar \kappa _1 \bar m_0\right),\nb\\
B_2&=& \frac{1}{2} \epsilon _{\text{Pl}} \bar y_0^{\sigma -2} (a+b-c) \left(\bar \kappa _0 \bar m_1+\bar \kappa _1 \bar m_0\right),\nb\\
B_3&=& -\frac{1}{2} \epsilon _{\text{Pl}} \bar y_0^{\sigma -2} \left(\bar \kappa _0 \bar m_1+\bar \kappa _1 \bar m_0\right) (a-b-c+d),\nb\\
B_4&=& -\frac{1}{2} \epsilon _{\text{Pl}} \bar y_0^{\sigma -2} \left(\bar \kappa _0 \bar m_1+\bar \kappa _1 \bar m_0\right) (b-c-d+e),\nb\\
B_5&=& -\frac{1}{2} \epsilon _{\text{Pl}} \bar y_0^{\sigma -2} \left(\bar \kappa _0 \bar m_1+\bar \kappa _1 \bar m_0\right) (c-d-e+f),\nb\\
B_6 &=& -\frac{1}{2} \epsilon _{\text{Pl}} \bar y_0^{\sigma -2} (d-e-f) \left(\bar \kappa _0 \bar m_1+\bar \kappa _1 \bar m_0\right),\nb\\
B_7&=&-\frac{1}{2} (e-f) \epsilon _{\text{Pl}} \bar y_0^{\sigma -2} \left(\bar \kappa _0 \bar m_1+\bar \kappa _1 \bar m_0\right),\nb\\
B_8&=&-\frac{1}{2} f \epsilon _{\text{Pl}} \bar y_0^{\sigma -2} \left(\bar \kappa _0 \bar m_1+\bar \kappa _1 \bar m_0\right).
\eqn

The error control function $\mathscr{H}(+\infty)$ can also be obtained by employing the expansions given in Eq.\ (\ref{slexpand}), and we find
\bqn\lb{Hinfty-inv}
\frac{\mathscr{H}(+\infty)}{\lambda} &\simeq & \frac{1}{6 \bar y_0}-\frac{(23+12 \ln2) \bar y_1}{72 \bar y_0^2}\nb\\
&&+\Bigg[\frac{\pi  a}{16}+\left(\frac{2}{3}-\frac{\pi }{16}\right) b+\left(\frac{15 \pi }{32}-\frac{2}{3}\right) c\nb\\
&&~~~~+\left(\frac{8}{3}-\frac{15 \pi }{32}\right) d+\left(\frac{175 \pi }{128}-\frac{8}{3}\right) e\nb\\
&&~~~~+\left(\frac{32}{5}-\frac{175 \pi }{128}\right) f-\frac{1}{12}\Bigg] \chi  \epsilon _{\text{Pl}} \bar y_0^{\sigma-1} \bar \kappa _0\nb\\
&&+\Bigg(\frac{a}{12}-\frac{2 d}{3}+\frac{2 e}{3}-\frac{8 f}{3}-\frac{c \pi }{16}+\frac{d \pi }{16}\nb\\
&&~~~~-\frac{15 e \pi }{32}+\frac{15 f \pi }{32}\Bigg)\bar m _0\epsilon_{\text{Pl}}\bar \kappa _0\bar y_0^{\sigma-3}\nb\\
&&+\Bigg[\left(\frac{23}{144}+\frac{\pi }{48}+\frac{\ln2}{12}\right)a-\left(\frac{1}{12}+\frac{\pi }{48}\right) b\nb\\
&&~~~~+\left(\frac{1}{12}-\frac{7 \pi}{48}+\frac{\pi\ln2}{16}\right) c\nb\\
&&~~~~+\left(\frac{7 \pi }{48}-\frac{1}{6}-\frac{2 \ln2}{3}-\frac{\pi  \ln2}{16} \right) d\nb\\
&&~~~~+\left(\frac{1}{6}-\frac{41 \pi}{64}+\frac{2 \ln2}{3}+\frac{15 \pi \ln2}{32}\right)e\nb\\
&&~~~~+\left(\frac{4}{9}+\frac{41 \pi }{64}-\frac{8 \ln2}{3}-\frac{15 \pi \ln2}{32}\right) f\Bigg]  \nb\\
&&~~~~~~~~~~\times\left(\bar \kappa _0 \bar m_1+\bar \kappa _1 \bar m_0\right)\epsilon_{\text{Pl}}\bar y_0^{\sigma-3}.\nb\\
\eqn
Once we get the integral of $\sqrt{g(y)}$ in Eq.~(\ref{int-g-inv}) and the error control function in Eq.~(\ref{Hinfty-inv}), from Eq.~(\ref{pw}) we can easily calculate the power spectra.

Now we turn to consider the corresponding spectral indices. In order to do this, we first specify the $k$-dependence of $\bar y_0(\eta_0)$, $\bar y_1(\eta_0)$, $\bar \epsilon_{\text{Pl}}(\eta_0)$, and $\bar m(\eta_0)$ through  $\eta_0 = \eta_0(k)$. From $g(y_0)=0$ we find
\bqn\lb{bary0}
\bar y_0 \simeq \bar \nu_0+\frac{1}{2} \epsilon_{\text{Pl}} \big(\bar m_0\bar \nu_0^{\sigma-1}-\chi \bar \nu_0^{\sigma+1}\big) \bar \kappa_0,
\eqn
and noticing $-k\eta_0 = \bar y_0(\eta_0)$ and $\epsilon_{\text{Pl}} \sim k^{-\sigma}$, we obtain
\bqn
\frac{d\ln(-\eta_0)}{d\ln k}& \simeq& -\left(1+\frac{\bar y_1}{\bar y_0}\right) \left[1+\frac{\sigma}{2} \epsilon_{\text{Pl}} \big(\bar m_0\bar \nu_0^{\sigma-2}-\chi \bar \nu_0^{\sigma}\big) \bar \kappa_0\right].\nb\\
\eqn
Then, using the above relation, the spectral indices are given by
\bqn
n-1 &\simeq &
3-2 \bar \nu_0+\left(\frac{1}{6\bar \nu _0^2}-\ln4\right)\bar \nu_1\nb\\
&&+\epsilon _{\text{Pl}} \left(\bar m_1 \bar \kappa _0+\bar m_0 \bar \kappa _1\right) \left(\Sigma_1 \bar \nu_0^{\sigma-3 }+\Sigma_2 \bar \nu_0^{\sigma-1}\right)\nb\\
&&+\epsilon _{\text{Pl}} \chi \bar \kappa_0 \left(\Sigma_3 \bar \nu_0^{\sigma-1 }+\Sigma_4 \bar \nu_0^{\sigma+1}\right)\nb\\
&&+\epsilon _{\text{Pl}} \chi \bar \kappa_1 \left(\Sigma_5 \bar \nu_0^{\sigma-1 }+\Sigma_6 \bar \nu_0^{\sigma+1}\right)\nb\\
&&+\epsilon _{\text{Pl}}\bar m_0 \bar \kappa _0\left(\Sigma_7 \bar \nu_0^{\sigma-3 }+\Sigma_8 \bar \nu_0^{\sigma-1}\right),
\eqn
where $\Sigma_i,\; i=1,\cdots 8$ depend on the value of $\sigma$ and are given in the Table I. Then the corresponding spectral index reads
\bqn
\alpha &\simeq &
2 \bar \nu _1-\sigma \epsilon _{\text{Pl}} \bar m_0 \bar \kappa _0 \left(\bar \nu _0^{\sigma-3} \Sigma _7+\bar \nu _0^{\sigma-1}  \Sigma _8\right)\nb\\
&&- \epsilon _{\text{Pl}} (m_1 \bar \kappa _0+\bar m_0 \bar \kappa _1) [\bar \nu _0^{\sigma-3} \left(\sigma  \Sigma _1+\Sigma _7\right)\nb\\
&&~~~~~~~~~~~~~~~~~~~~~~~~~~~~+\bar \nu _0^{\sigma -1} \left(\sigma  \Sigma _2+\Sigma _8\right)\Big]\nb\\
&&-\sigma \chi  \epsilon _{\text{Pl}} \bar \kappa _0 \left(\bar \nu _0^{\sigma-1} \Sigma _3+\bar \nu _0^{\sigma+1 }  \Sigma _4\right).
\eqn

\section*{Appendix C: Slow-roll expansions of $\nu(\eta)$, $c_s(\eta)$, $m(\eta)$,  $\kappa(\eta)$,  and their derivatives}
\renewcommand{\theequation}{C.\arabic{equation}} \setcounter{equation}{0}

\subsection{Expansions with the holonomy corrections}

Let us first consider the scalar perturbation with the holonomy corrections. Using the expression of $z''/z$ in Eq. (\ref{zsh}), it is easy to find that $\nu(\eta)$ for scalar perturbations reads
\bqn\lb{nus}
\nu^s(\eta)&\simeq& \frac{3}{2}+\epsilon _1+\frac{\epsilon _2}{2}+\epsilon _1^2+\frac{11 \epsilon _2 \epsilon _1}{6}+\frac{\epsilon _2 \epsilon _3}{6}-2 \epsilon _1 \delta _H\nb\\
&&-\frac{2\epsilon _1^2 \delta _H}{3} -6 \epsilon _1 \delta _H^2-\frac{2\epsilon _2 \epsilon _1 \delta _H}{3} +\epsilon _1^3+\frac{77}{18} \epsilon _2 \epsilon _1^2\nb\\
&&+\frac{17}{9} \epsilon _2^2 \epsilon _1+\frac{14}{9} \epsilon _2 \epsilon _3 \epsilon _1-\frac{\epsilon _2^2 \epsilon _3}{18} .
\eqn
Consideration of the derivatives of $\nu^s(\eta)$ with respect to $\ln{(-\eta)}$ yields
\bqn\lb{nu1s}
\nu_{1}^s&\simeq& -\epsilon _1 \epsilon _2-\frac{\epsilon _3 \epsilon _2}{2}-4 \epsilon _1^2 \delta _H+2 \epsilon _2 \epsilon _1 \delta _H-3 \epsilon _2 \epsilon _1^2\nb\\
&&-\frac{11}{6} \epsilon _2^2 \epsilon _1-\frac{7\epsilon _2 \epsilon _3}{3}  \epsilon _1-\frac{\epsilon _2 \epsilon _3^2}{6} -\frac{\epsilon _2 \epsilon _3 \epsilon _4}{6} \nb\\
&&-\frac{16}{3}  \epsilon _1^3 \delta _H-28 \epsilon _1^2 \delta _H^2+2 \epsilon _2 \epsilon _1^2 \delta _H+\frac{2}{3} \epsilon _2^2 \epsilon _1 \delta _H\nb\\
&&+6 \epsilon _2 \epsilon _1 \delta _H^2+\frac{2}{3} \epsilon _2 \epsilon _3 \epsilon _1 \delta _H-6 \epsilon _2 \epsilon _1^3-\frac{205}{18} \epsilon _2^2 \epsilon _1^2\nb\\
&&-\frac{119}{18} \epsilon _2 \epsilon _3 \epsilon _1^2-\frac{17}{9} \epsilon _2^3 \epsilon _1-\frac{31}{18} \epsilon _2 \epsilon _3^2 \epsilon _1-\frac{35}{6} \epsilon _2^2 \epsilon _3 \epsilon _1\nb\\
&&-\frac{31}{18} \epsilon _2 \epsilon _3 \epsilon _4 \epsilon _1+\frac{\epsilon _2^2 \epsilon _3^2}{9} +\frac{\epsilon _2^2 \epsilon _3 \epsilon _4}{18} ,
\eqn
\bqn\lb{nu2s}
\nu_{2}^s&\simeq& \epsilon _1 \epsilon _2^2+\frac{\epsilon _3^2 \epsilon _2}{2} +\epsilon _1 \epsilon _3 \epsilon _2+\frac{ \epsilon _3 \epsilon _4 \epsilon _2}{2}-8 \epsilon _1^3 \delta _H\nb\\
&&+12 \epsilon _2 \epsilon _1^2 \delta _H-2 \epsilon _2^2 \epsilon _1 \delta _H-2 \epsilon _2 \epsilon _3 \epsilon _1 \delta _H+7 \epsilon _2^2 \epsilon _1^2\nb\\
&&+4 \epsilon _2 \epsilon _3 \epsilon _1^2+\frac{11}{6} \epsilon _2^3 \epsilon _1+\frac{17}{6} \epsilon _2 \epsilon _3^2 \epsilon _1+6 \epsilon _2^2 \epsilon _3 \epsilon _1\nb\\
&&+\frac{17}{6} \epsilon _2 \epsilon _3 \epsilon _4 \epsilon _1+\frac{\epsilon _2 \epsilon _3^3}{6} +\frac{ \epsilon _2 \epsilon _3 \epsilon _4^2}{6}+\frac{\epsilon _2 \epsilon _3^2 \epsilon _4}{2} \nb\\
&&+\frac{\epsilon _2 \epsilon _3 \epsilon _4 \epsilon _5}{6} ,
\eqn
and
\bqn\lb{nu3s}
\nu_{3}^s&\simeq& -\epsilon _1 \epsilon _2^3-3 \epsilon _1 \epsilon _3 \epsilon _2^2-\frac{\epsilon _3^3 \epsilon _2}{2} -\epsilon _1 \epsilon _3^2 \epsilon _2-\frac{\epsilon _3 \epsilon _4^2 \epsilon _2}{2} \nb\\
&&-\frac{3}{2} \epsilon _3^2 \epsilon _4 \epsilon _2-\epsilon _1 \epsilon _3 \epsilon _4 \epsilon _2-\frac{1}{2} \epsilon _3 \epsilon _4 \epsilon _5 \epsilon _2.
\eqn

For the tensor perturbations, using Eq. (\ref{zth}), we find
\bqn\lb{nut}
\nu^t(\eta) &\simeq &  \frac{3}{2}+\epsilon _1+\epsilon_1^2+\frac{4 \epsilon _1 \epsilon_2}{3}-2 \delta _H \epsilon _1-6 \delta _H^2 \epsilon _1+\epsilon _1^3\nb\\
&&-\frac{2}{3} \delta _H \epsilon _1^2-\frac{2}{3} \delta _H \epsilon _1 \epsilon _2+\frac{34}{9} \epsilon _1^2 \epsilon _2+\frac{4}{3} \epsilon _1 \epsilon _2^2+\frac{4}{3} \epsilon _1 \epsilon _2 \epsilon _3,\nb\\
\eqn
\bqn\lb{nu1t}
\nu_{1}^t &\simeq & -\epsilon _1 \epsilon _2-4 \delta _H \epsilon _1^2+2 \delta _H \epsilon _1 \epsilon _2-3 \epsilon _1^2 \epsilon_2-\frac{4}{3} \epsilon _1 \epsilon _2^2\nb\\
&&-\frac{4}{3} \epsilon _1 \epsilon _2 \epsilon_3+28 \delta _H^2 \epsilon _1^2+\frac{16}{3} \delta _H \epsilon _1^3-6 \delta _H^2 \epsilon _1 \epsilon _2\nb\\
&&-2 \delta _H \epsilon _1^2 \epsilon _2+6 \epsilon _1^3 \epsilon _2-\frac{2}{3} \delta _H \epsilon _1 \epsilon _2^2+\frac{89}{9} \epsilon _1^2 \epsilon _2^2\nb\\
&&+\frac{4}{3} \epsilon _1 \epsilon _2^3-\frac{2}{3} \delta _H \epsilon _1 \epsilon _2 \epsilon _3+\frac{46}{9} \epsilon _1^2 \epsilon _2 \epsilon _3+4 \epsilon _1 \epsilon _2^2 \epsilon _3\nb\\
&&+\frac{4}{3} \epsilon _1 \epsilon _2 \epsilon _3^2+\frac{4}{3} \epsilon _1 \epsilon _2 \epsilon _3 \epsilon _4,
\eqn
\bqn\lb{nu2t}
\nu_{2}^t &\simeq & \epsilon _1 \epsilon _2^2+\epsilon _1 \epsilon _2 \epsilon _3-8 \delta _H \epsilon _1^3+12 \delta _H \epsilon _1^2 \epsilon _2-2 \delta _H \epsilon _1 \epsilon _2^2\nb\\
&&+7 \epsilon _1^2 \epsilon _2^2+\frac{4}{3} \epsilon _1 \epsilon _2^3-2 \delta _H \epsilon _1 \epsilon _2 \epsilon _3+4 \epsilon _1^2 \epsilon _2 \epsilon _3\nb\\
&&+4 \epsilon _1 \epsilon _2^2 \epsilon _3+\frac{4}{3} \epsilon _1 \epsilon _2 \epsilon _3^2+\frac{4}{3} \epsilon _1 \epsilon _2 \epsilon _3 \epsilon _4,
\eqn
and
\bqn\lb{nu3t}
\nu_{3}^t&\simeq & -\epsilon _1 \epsilon _2^3-3 \epsilon _1 \epsilon _3 \epsilon _2^2-\epsilon _1 \epsilon _3^2 \epsilon _2-\epsilon _1 \epsilon _3 \epsilon _4 \epsilon _2.\nb\\
\eqn

Now we turn to consider $c_s(\eta)$. Expanding it in terms of $\delta_H$, we observe that
\bqn
c_s(\eta) &\equiv& \sqrt{1-2 \delta_H}\simeq 1-\delta_H-\frac{\delta _H^3}{2}-\frac{\delta _H^2}{2}.
\eqn
Similar to $\nu^s(\eta)$, the derivatives of $c_s(\eta)$ with respect to $\ln{(-\eta)}$ are given by
\bqn
c_1(\eta)&\simeq & -2 \epsilon _1 \delta _H-2 \epsilon _1^2 \delta _H-4 \epsilon _1 \delta _H^2-9 \epsilon _1 \delta _H^3-4 \epsilon _1^2 \delta _H^2\nb\\
&&-2 \epsilon _1^2 \epsilon _2 \delta _H-2 \epsilon _1^3 \delta _H,
\eqn
\bqn
c_2(\eta)&\simeq & -4 \epsilon _1^2 \delta _H+2 \epsilon _2 \epsilon _1 \delta _H-8 \epsilon _1^3 \delta _H-20 \epsilon _1^2 \delta _H^2\nb\\
&&+6 \epsilon _2 \epsilon _1^2 \delta _H+4 \epsilon _2 \epsilon _1 \delta _H^2,
\eqn
and
\bqn
c_3(\eta)&\simeq&-8 \epsilon _1^3 \delta _H+12 \epsilon _2 \epsilon _1^2 \delta _H-2 \epsilon _2^2 \epsilon _1 \delta _H-2 \epsilon _2 \epsilon _3 \epsilon _1 \delta _H.\nb\\
\eqn
Note that for both scalar and tensor perturbations, the effective sound speed $c_s(\eta)$ takes the same form. Thus in the following, we are not going to distinguish them.

\subsection{Expansions with inverse-volume corrections}

Now let us turn to consider the slow-roll expansions of $\nu$, $m(\eta)$ and $\kappa(\eta)$.  For $\nu(\eta)$ and its derivatives, it is worth to note that when one considers the inverse-volume corrections, $\nu(\eta)$ and its derivatives with respect to $\ln(-\eta)$, i.e., $\nu_0,\;\nu_1,\;\nu_2,\;\text{and}\;\nu_3$ all take the same form as those given in general relativity. Thus, they can be directly obtained from Eqs. (\ref{nus})-(\ref{nu3s}) for the scalar perturbations and from Eqs. (\ref{nut})-(\ref{nu3t}) for the tensor perturbations by taking the holonomy parameter $\delta_H=0$. For the scalar perturbations, on the other hand, the function $m(\eta)$ reads
\bqn
m^s(\eta)&=&\frac{\sigma^2\left(3- \sigma\right) \alpha _0}{4 \epsilon_1}\nb\\
&&+\left(\frac{3 \sigma }{2}-\frac{\sigma ^2}{4}-\frac{\sigma ^3}{12}\right) \vartheta_0+\left(\frac{5 \sigma ^2}{4}-\frac{\sigma ^3}{2}\right) \alpha_0\nb\\
&&+\frac{\sigma  \alpha _0 (\sigma -3)\epsilon _2}{4 \epsilon_1}+\frac{\sigma  \alpha _0\epsilon _2 \epsilon _3}{4 \epsilon _1} .
\eqn
Note that at the turning point we write $m(\eta_0)=\bar m_0$. Now consideration of the derivatives of $m(\eta)$ with respect to $\ln(-\eta)$ yields
\bqn
m_1^s &=&\frac{\sigma ^2 \alpha _0 (3-\sigma ) \epsilon _2}{4 \epsilon _1} .
\eqn
Similarly, for tensor perturbations, we have
\bqn
m^t(\eta) &\simeq& \frac{3 \sigma  \alpha _0}{2}-\frac{\sigma ^2 \alpha _0}{2},
\eqn
while up to the second-order in the slow-roll parameters we have $m_2^t \simeq 0$ and $m_3^t \simeq 0$. Note that we also write $m^t(\eta_0)$ as $\bar m_0$.

To get the slow-roll expansions of  the inflationary observables, we also need the slow-roll expansions of $\kappa(\eta)$ and its derivatives, which are given by
\bqn
\kappa(\eta)&=& H^\sigma \Big(1-\sigma  \epsilon _1\Big),
\eqn
\bqn
\kappa_1&=& \sigma H^\sigma \epsilon _1 .\nb\\
\eqn
Note that at the turning point we write $\kappa(\eta_0)=\bar \kappa_0$.

\begin{widetext}

\begin{table}
\caption{\label{tab:table1} Values of Coefficients $\Sigma_i (i=1,\dots, 8)$ for different values of $\sigma$.}
\begin{ruledtabular}
\begin{tabular}{ccccccc}
$\sigma$ &1 & 2 & 3 & 4&5 & 6 \\
\hline
$\Sigma_1$ & $-\frac{\pi}{48}$ &$\frac{1}{6}$ &$ \frac{(8-3\ln2)\pi}{16} $&  $\frac{4+8\ln2}{3}$ & $\frac{5(47-30\ln2)\pi}{64}$ &$16\ln2$ \\
\hline
$\Sigma_2$ & $\frac{(\ln2-1)\pi}{2} $ & $1-2\ln2$ & $\frac{(6\ln2-5)\pi}{8} $&$\frac{14-24\ln2}{9} $ & $\frac{(60\ln2-47)\pi}{64}$ & $\frac{4(37-60\ln2)}{75}$\\
\hline
$\Sigma_3$ & $-\frac{\pi }{16}$ & $-\frac{4}{3}$ & $-\frac{45 \pi}{32}$ & $-\frac{32}{3}$ & $-\frac{875 \pi}{128}$ & $-\frac{192}{5}$ \\
\hline
$\Sigma_4$ & $\frac{\pi }{4} $ & $\frac{4}{3}$ & $\frac{9 \pi }{16}$ & $\frac{32}{15} $ & $\frac{25 \pi }{32}$ & $\frac{96}{35}$\\
\hline
$\Sigma_5$ & $-\frac{\pi }{16}-\frac{23}{144}-\frac{\ln2}{12}$ & $-\frac{71}{72}-\frac{\ln2}{6} $ & $-\frac{15 \pi }{32}-\frac{23}{48}-\frac{\ln2}{4}$ &$ -\frac{119}{36}-\frac{\ln2}{3}$ & $-\frac{175 \pi }{128}-\frac{115}{144}-\frac{5 \ln2}{12} $ & $ -\frac{883}{120}-\frac{\ln2}{2}$\\
\hline
$\Sigma_6$ & $ -\frac{\pi }{16}-\frac{23}{144}-\frac{\ln2}{12}$ & $-\frac{71}{72}-\frac{\ln2}{6}$ & $-\frac{15 \pi }{32}-\frac{23}{48}-\frac{\ln2}{4}$ & $-\frac{119}{36}-\frac{\ln2}{3}$ & $-\frac{175 \pi }{128}-\frac{115}{144}-\frac{5 \ln2}{12}  $ & $-\frac{883}{120}-\frac{\ln2}{2}$ \\
\hline
$\Sigma_7$ & $0$ & $0$ & $\frac{3 \pi }{16}$ &$\frac{8}{3} $ &$\frac{75 \pi }{32}$ &$16$\\
\hline
$\Sigma_8$ & $-\frac{\pi }{2}$ & $-2$ &$-\frac{3 \pi}{4}$ &$ -\frac{8}{3}$ & $-\frac{15 \pi}{16}$ & $-\frac{16}{5}$
\end{tabular}
\end{ruledtabular}
\end{table}

\begin{table}
\centering
\caption{\label{tab:table3} Values of Coefficients $\mathcal{\bar Q}_{i}^{(s,t)}$, $\mathcal{\bar K}_{i}^{(s,t)}$, $\mathcal{\bar L}_i^{(s,t)} (i=-1,0,1)$ for different values of $\sigma$.}
\begin{ruledtabular}
\begin{tabular}{ccccccc}
$\sigma$ &1 & 2 & 3 & 4&5 & 6 \\
\hline
$\frac{\mathcal{\bar Q}_{-1}^{(s)}}{\alpha_0}$ & $\frac{\pi }{6}$&$\frac{2}{3} $&$0$&$-\frac{1616 }{1629}$&$\frac{475 \pi }{2896} $&$\frac{10512}{905} $ \\
\hline
$\frac{\mathcal{\bar Q}_0^{(s)} }{\alpha_0}$ & $-\frac{2}{9}-\frac{45401 \pi }{52128}+\frac{\pi \ln2}{6} $ & $\frac{8\ln2}{3} -\frac{14647 }{3258}$ & $\frac{513 \pi}{11584}$ & $\frac{82702 }{8145}-\frac{3232\ln2}{543}  $ & $\frac{50}{9}+\frac{743995 \pi }{139008}-\frac{1425 \pi  \ln2}{2896}$ & $\frac{84096\ln2}{905} -\frac{715864 }{31675}$\\
\hline
$\frac{\mathcal{\bar Q}_1^{(s)}}{\alpha_0}$ & $\frac{176 \pi }{1629}-\frac{1}{9}$ & $\frac{4\ln2}{3}-\frac{985}{1629}$ & $0$ & $\frac{5732}{4887}-\frac{3232 \ln2}{1629}$ & $\frac{25 }{9}+\frac{17165 \pi }{34752}$ & $\frac{58418 }{4525}+\frac{21024\ln2}{905}  $\\
\hline
$\frac{\mathcal{\bar K}_{-1}^{(s)}}{\alpha_0}$ & $-\frac{\pi }{6}$ & $-\frac{4}{3}$ & $0$ & $\frac{320}{81}$ & $-\frac{125 \pi}{144} $ & $-\frac{352}{5}$ \\
\hline
$\frac{\mathcal{\bar K}_{0}^{(s)}}{\alpha_0}$& $ \frac{20 \pi }{81}+\frac{\pi  \ln2}{6} $ & $\frac{251}{81}-\frac{8 \ln2}{3}$ & $-\frac{9 \pi}{64}$ & $\frac{1280 \ln2}{81}-\frac{22696}{1215}$ & $\frac{625 \pi  \ln2}{144}-\frac{104075 \pi }{3456}$ & $-\frac{26888}{525}-\frac{2112 \ln2}{5} $\\
\hline
$\frac{\mathcal{\bar K}_{1}^{(s)}}{\alpha_0}$ & $\frac{\pi  \ln2}{6} -\frac{109 \pi }{324}$ & $-\frac{32}{81}-\frac{4 \ln2}{3} $ & $0$ & $\frac{320}{243}+\frac{320 \ln2}{81}$ & $\frac{125\pi  \ln2}{144} -\frac{5225 \pi }{1728}$ & $-\frac{5944}{75}-\frac{352 \ln2}{5} $\\
\hline
$\frac{\mathcal{\bar K}_{2}^{(s)}}{\alpha_0}$ & $\frac{10165 \pi }{15552}-\frac{ \pi  \ln2}{12}$ & $\frac{524}{243}-2 \ln2$ & $\frac{27\pi  \ln2}{64} -\frac{197 \pi }{256}$ & $\frac{58976 \ln2}{1215}-\frac{101072}{2025}$ & $-\frac{13175 \pi }{2592}-\frac{95875 \pi  \ln2}{3456}$ & $\frac{4432 \ln2}{35}-\frac{913732}{3675}$\\
\hline
$\frac{\mathcal{L}^{(s)}_{-1}}{\alpha_0}$ & $\frac{\pi }{6}$ & $\frac{8}{3}$ & $0$ & $-\frac{1280}{81}$ & $\frac{625 \pi }{144}$ & $\frac{2112}{5}$\\
\hline
$\frac{\mathcal{L}^{(s)}_{0}}{\alpha_0}$ & $-\frac{13 \pi }{162}-\frac{\pi  \ln2}{6} $ & $\frac{16 \ln2}{3}-\frac{286}{81}$ & $\frac{27 \pi }{64}$ & $\frac{71584}{1215}-\frac{5120 \ln2}{81}$ & $\frac{535375 \pi }{3456}-\frac{3125\pi  \ln2}{144} $ & $\frac{127696}{175}+\frac{12672 \ln2}{5}$\\
\hline
$\frac{\mathcal{L}^{(s)}_{1}}{\alpha_0}$ & $\frac{163 \pi }{324}-\frac{\pi  \ln2}{6} $ & $\frac{172}{81}+\frac{8 \ln2}{3}$ & $0$ & $-\frac{2240}{243}-\frac{1280 \ln2}{81} $ & $\frac{27625 \pi }{1728}-\frac{625\pi  \ln2}{144} $ & $\frac{13648}{25}+\frac{2112 \ln2}{5}$\\
\hline
$\frac{\mathcal{\bar Q}_0^{(t)}}{\alpha_0}$ & $-\frac{725 \pi  }{2172}$ & $-\frac{244}{543} $ & $0$ & $\frac{11728}{8145}$ & $\frac{8165 \pi }{5792}$ & $\frac{13920}{1267}$\\
\hline
$\frac{\mathcal{\bar K}_0^{(t)}}{\alpha_0}$ & $\frac{\pi }{3}$ & $\frac{8}{9}$ & $0$ & $-\frac{2368}{405}$ & $-\frac{1025 \pi}{144} $ & $-\frac{6976}{105}$\\
\hline
$\frac{\mathcal{\bar L}_0^{(t)}}{\alpha_0}$ & $-\frac{\pi }{3}$ & $-\frac{16}{9}$ & $0$ & $\frac{9472}{405}$ & $\frac{5125 \pi }{144}$ & $\frac{13952}{35}$ \\
\end{tabular}
\end{ruledtabular}
\end{table}

\end{widetext}


\baselineskip=12truept

\begin{thebibliography}{99}




\bibitem{Guth}
A. Guth, %
Phys. Rev. D{\bf 23}, 347 (1981); %
A.A. Starobinsky, %
Phys. Lett. B {\bf 91}, 99 (1980); %
K. Sato, %
Mon. Not. R. Astron. Soc. {\bf 195}, 467 (1981).

\bibitem{InfGR}
D. Baumann, %
arXiv:0907.5424.

\bibitem{WMAP}
E. Komatsu {\em et al.} (WMAP Collaboration), %
Astrophys. J. Suppl. Ser. {\bf 192}, 18 (2011); %
D. Larson {\em et al.} (WMAP Collaboration),  %
{\em ibid.},  {\bf 192}, 16 (2011).

\bibitem{PLANCK}
P. Ade {\em et al.} (PLANCK Collaboration), %
arXiv:1303.5082.



\bibitem{BICEP2}
P.A.R. Ade et al. (BICEP2 Collaboration), %
\PRL {\bf 112}, 241101 (2014).

\bibitem{Planck-intermediate}
P.A.R. Ade, {\em et al.} (BICEP2/Keck and Planck Collaborations), arXiv:1502.00612; R. Adam, {\em et al.}  (Planck Collaborations), arXiv:1502.01582.
  
\bibitem{Mort2014}
M.J. Mortonson and U. Seljak, %
\JCAP 10 (2014) 035; 
R. Flauger, J.C. Hill and D.N. Spergel, %
\JCAP \; 1408 (2014) 039. 

\bibitem{DB}  
C.P. Burgess, M. Cicoli, and F. Quevedo,
JCAP {\bf 11} (2013) 003; 
R.H. Brandenberger and J. Martin, Class. Quantum. Grav. {\bf 30} (2013) 113001.

\bibitem{Bojowald2001}
M. Bojowald, %
Phys. Rev. Lett. {\bf 86}, 5227 (2001).

\bibitem{Ashtekar2006}
A. Ashtekar, T. Pawlowski, and P. Singh, %
Phys. Rev. Lett. {\bf 96}, 141301 (2006);
Phys. Rev. D{\bf 73}, 124038 (2006); %
Phys. Rev. D{\bf 74}, 084003 (2006); %
A. Ashtekar, A. Corichi, and P. Singh, %
Phys. Rev. D{\bf 77}, 024046 (2008).

\bibitem{QGEs}  M. Bojowald, Rep. Prog. Phys. {\bf 78} (2015) 023901;
A.  Ashtekar and A. Barrau,  arXiv:1504.07559.



\bibitem{Mielczarek2008}
J. Mielczarek, %
J. Cosmol. Astropart. Phys. {\bf 11} (2008) 011; %
Phys. Rev. D{\bf 79}, 123520 (2009); %
J. Mielczarek, T. Cailleteau, J. Grain,  and A. Barrau, %
Phys. Rev. D{\bf 81}, 104049 (2010).

\bibitem{Grain2009PRL}
J. Grain and A. Barrau, %
Phys. Rev. Lett. {\bf 102}, 081301 (2009).

\bibitem{Grain2010PRD}
J. Grain, A. Barrau, T. Cailleteau, and J. Mielczarek, %
Phys. Rev. D{\bf 82}, 123520 (2010).

\bibitem{vector_holonomy}
Y. Li and J.-Y Zhu, %
Class. Quantum Grav. {\bf 28}, 045007 (2011); %
J. Mielczarek, T. Cailleteau, A. Barrau and J. Grain, %
Class. Quantum Grav. {\bf 29}, 085009 (2012).


\bibitem{scalar}
T. Cailleteau, J. Mielczarek,  A. Barrau and J. Grain, %
Class. Quantum Grav. {\bf 29}, 095010 (2012).

\bibitem{scalar2}
T. Cailleteau, A. Barrau, F. Vidotto, and J, Grain, %
Phys. Rev. D{\bf 86}, 087301 (2012).


\bibitem{loop_corrections}
A. Barrau, T. Cailleteau, J. Grain, and J. Mielczarek, %
arXiv: 1309.6896. 


\bibitem{Bojowald2008}
M. Bojowald and G.M. Hossain, %
Phys. Rev. D{\bf 78}, 063547 (2008).

\bibitem{Bojowald2009}
M. Bojowald, G.M. Hossain, M. Kagan, and S. Shankaranarayanan, %
Phys. Rev. D{\bf 79}, 043505 (2009); 
D{\bf 82}, 109903 (E) (2010).

\bibitem{Bojowald2007}
M. Bojowald and G.M. Hossain, %
Classical Quantum Gravity {\bf 24}, 4801 (2007).

\bibitem{Bojowald2008b}
M. Bojowald and G.M. Hossain, %
Phys. Rev. D{\bf 77}, 023508 (2008).

\bibitem{Bojowald2011}
M. Bojowald and G. Calcagni, %
JCAP 03 (2011) 032.


\bibitem{Bojowald2011b}
M. Bojowald, G. Calcagni, and S. Tsujikawa, %
Phys. Rev. Lett. {\bf 107}, 211302 (2011);
M. Bojowald, G. Calcagni, and S. Tsujikawa, %
\JCAP 11 (2011) 046.


\bibitem{Mielczarek2014}
J. Mielczarek, %
JCAP 03 (2014) 048. 



\bibitem{Cai2012}
L.-F. Li, R.-G. Cai, Z.-K. Guo, and B. Hu, %
Phys. Rev. D{\bf 86}, 044020 (2012).

\bibitem{Zhu1}
T. Zhu, A. Wang, G. Cleaver, K. Kirsten, and Q. Sheng, %
Int. J. Mod. Phys. A{\bf 29},  1450142 (2014). 

\bibitem{Zhu2}
T. Zhu, A. Wang, G. Cleaver, K. Kirsten, and Q. Sheng, %
Phys. Rev. D{\bf 89}, 043507 (2014); 
T. Zhu and A. Wang, %
Phys. Rev. D{\bf 90}, 027304 (2014). 

\bibitem{Uniform3}
T. Zhu, A. Wang, G. Cleaver, K. Kirsten, and Q. Sheng, %
Phys. Rev. D{\bf 90}, 063503 (2014). 

\bibitem{Uniform4}
T. Zhu, A. Wang, G. Cleaver, K. Kirsten, and Q. Sheng, %
Phys. Rev. D{\bf 90}, 103517 (2014).

\bibitem{uniformPRL}
S. Habib, K. Heitmann, G. Jungman, and C. Molina-Paris, %
Phys. Rev. Lett. {\bf 89}, 281301 (2002); %
S. Habib, A. Heinen, K. Heitmann, G. Jungman, and C. Molina-Paris, %
Phys. Rev. D{\bf 70}, 083507 (2004); %
S. Habib, A.Heinen, K. Heitmann, and G. Jungman, %
Phys. Rev. D{\bf 71}, 043518 (2005).


\bibitem{Zhu3}
T. Zhu, A. Wang, G. Cleaver, K. Kirsten, and Q. Sheng, %
Astrophys. J. Lett. {\bf 807}, L17 (2015). 



\bibitem{slow-roll-holonomy}
J. Mielczarek, %
Phys. Rev. D{\bf 81}, 063501 (2010).

\bibitem{minisuper}
G. Calcagni and G.M. Hossain, %
Adv. Sci. Lett. {\bf 2}, 184 (2009); %
A. Ashtekar, T. Pawlowshi,  and P. Singh, %
Phys. Rev. D{\bf 74}, 084003 (2006).



\bibitem{Bojowald2009c}
M. Bojowald, G.M. Hossain, M. Kagan, and S. Shankaranarayanan, %
Phys. Rev. D{\bf 79}, 043505 (2009).


\bibitem{JM}     S.E. Joras and G. Marozzi, Phys. Rev. D{\bf 79}, 023514 (2009);
A. Ashoorioon, D. Chialva and U. Danielsson,   JCAP  {\bf 06}, 034 (2011).


 \bibitem{phi2}
P. Creminelli, D.L. Nacir, M. Simonovic, G. Trevisan, and M. Zaldarriaga, %
\PRL \; {\bf 112}, 241303 (2014).

\bibitem{cosmo}  E. Komatsu {\em et al.} (WMAP Collaboration), Astrophys. J. Suppl.  {\bf 192}, 18 (2011);
D. Larson {\em et al.} (WMAP Collaboration),  {\em ibid.},  {\bf 192}, 16 (2011);
  P.A.R. Ade {\em et al.} (PLANCK Collaboration), A$\&$A, {\bf 571}, A16 (2014);
   P.A.R. Ade, {\em et al.} (Planck Collaborations), arXiv:1502.02114.


 \bibitem{S4-CMB} K.N. Abazajian {\em et al.}, Astropart. Phys. {\bf 63}, 55 (2015) [arXiv:1309.5381].


\bibitem{COSMOMC}
http://cosmologist.info/cosmomc/; Y.-G. Gong, Q. Wu, and  A. Wang, 	Astrophys. J. {\bf 681},  27 (2008).

\bibitem{Planck2013}
P. A. R. Ade (Planck Collaboration), Astron. Astrophys. {\bf 571}  (2014) A16. 


\bibitem{BAO2013}
L. Anderson {\em et al.}, %
Mon. Not. R. Astron. Soc. {\bf 427}, 3435 (2013).

\bibitem{SN}
A. Conley, J. Guy, M. Sullivan, N. Regnault, P. Astier, C. Balland, S. Basa and R. G. Carlberg {\em et al.}, %
Astrophys. J. Suppl. {\bf 192}, 1 (2011).

\bibitem{axion}
E. Silverstein and A. Westphal, %
Phys. Rev. D{\bf 78}, 106003 (2008);
L. McAllister, E. Silverstein, and A. Westphal, %
Phys. Rev. D{\bf 82}, 046003 (2010).

\bibitem{Olver1974}
F.W.J. Olver, {\em Asymptotics and Special functions}, (AKP Classics, Wellesley, MA 1997).
\end{thebibliography}
\end{document}